\DocumentMetadata{}
\documentclass[sigconf]{acmart}

\newcommand{\eg}[0]{\textit{e.g., }}
\newcommand{\ie}[0]{\textit{i.e., }}

\newcommand{\ours}{\texttt{\tool{}}}

\usepackage{comment}
\usepackage{stfloats}
\usepackage{microtype}
\usepackage{enumitem}

\AtBeginDocument{%
  \providecommand\BibTeX{{%
    \normalfont B\kern-0.5em{\scshape i\kern-0.25em b}\kern-0.8em\TeX}}}

\usepackage[autostyle,german=guillemets]{csquotes}
\usepackage[utf8]{inputenc}
\makeatletter
\usepackage{hyperref,quoting}
\usepackage{url}
\quotingsetup{vskip=0pt,font={itshape,raggedright},rightmargin=0pt}
\usepackage{tikz}
\usepackage{cuted}
\usepackage{capt-of}
\usepackage{subcaption}
\usepackage{fancyvrb}
\usepackage{gensymb}
\usepackage{float}
\usepackage{stfloats}
\usepackage{tikz}
\usepackage{xcolor}

\usepackage{dis26-156}

\newcommand{\tool}[1]{\texttt{LearnMate$^2$}}

\copyrightyear{2026}
\acmYear{2026}
\acmConference[DIS '26]{Designing Interactive Systems Conference}{June 13--17, 2026}{Singapore, Singapore}
\acmBooktitle{Designing Interactive Systems Conference (DIS '26), June 13--17, 2026, Singapore, Singapore}
\acmDOI{10.1145/3800645.3812972}
\acmISBN{979-8-4007-2563-0/2026/06}

\begin{document}

%%
%% The "title" command has an optional parameter,
%% allowing the author to define a "short title" to be used in page headers.

% design evalution of the personalize, appdacive 11m-based suppoart learning sytem
\title{\texttt{\tool{}}: Design and Evaluation of an LLM-powered Personalized and Adaptive Support System for Online Learning}
%Design and Evaluation of an LLM-Based Personalized and Adaptive Learning Support System

%verp
%\tool{}
%surePlan

%Integrating Formal Verification for LLM-Based Planning Tools

% user-centric verification for llm-based planning tools

%Applying formal verification to LLMs for enhanced user control
%Verifying LLMs Enhancing User Control in Complex Task Planning with Temporal Constraints
%VERL: Integrating Formal Verification and User Control for Effective LLM-Based Planning Tools

%%
%% The "author" command and its associated commands are used to define
%% the authors and their affiliations.
%% Of note is the shared affiliation of the first two authors, and the
%% "authornote" and "authornotemark" commands
%% used to denote shared contribution to the research.

\author{Xinyu Jessica Wang}
\orcid{0009-0002-5519-8432}
\affiliation{%
  \institution{Department of Computer Sciences University of Wisconsin--Madison}
  \country{Madison, Wisconsin, USA}
}
\email{xwang2775@wisc.edu}

\author{Christine P Lee}
\orcid{0000-0003-0991-8072}
\affiliation{%
  \institution{Department of Computer Sciences University of Wisconsin--Madison}
  \country{Madison, Wisconsin, USA}
}
\email{cplee5@cs.wisc.edu}

\author{Bilge Mutlu}
\orcid{0000-0002-9456-1495}
\affiliation{%
  \institution{Department of Computer Sciences University of Wisconsin--Madison}
  \country{Madison, Wisconsin, USA}
}
\email{bilge@cs.wisc.edu}

\renewcommand{\shortauthors}{}

\begin{abstract}

Personalization is crucial for effective learning, yet online learning, designed for widespread availability and open access, lacks personalized guidance. Recent advancements in large language models (LLMs) offer opportunities to bridge this gap. We explore how LLM-driven tools may be designed to support personalized and adaptive learning and examine how they shape user experience and learning outcomes. We iteratively designed \tool{} to support online learning by providing personalized study plans, real-time contextual assistance, and adaptive learning activities. A preliminary study ($n=24$) assessed the effectiveness and usability of \tool{} and informed refinements in our system, which we then evaluated ($n = 16$) against a combination of a state-of-the-art online learning platform and an LLM for learning support. Results indicate that \tool{} advances AI pedagogy by improving both learning outcomes and user experience compared to existing online learning and support tools. This work advances our understanding of the design space of personalized, AI-driven educational tools and their potential impact on user experience.

\end{abstract}

\begin{teaserfigure}
  \includegraphics[width=\textwidth]{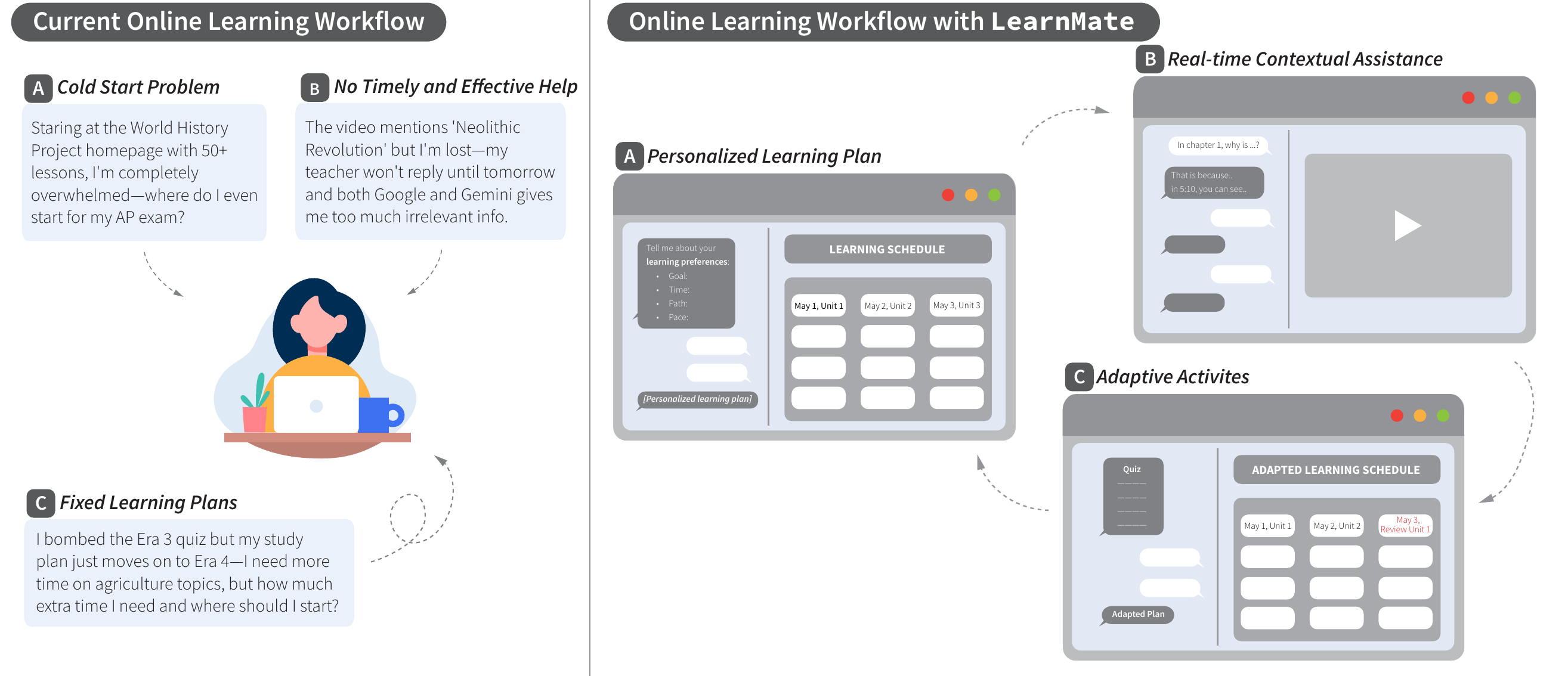}
  \Description[LearnMate² system overview]{A two-panel diagram. The left panel, labeled "Current Online Learning Workflow," shows three challenges: (A) Cold Start Problem, illustrated by a learner overwhelmed by 50+ lessons not knowing where to start; (B) No Timely and Effective Help, where a learner cannot get a contextual answer about the Neolithic Revolution; and (C) Fixed Learning Plans, where a learner who failed a quiz has no way to adjust their plan. The right panel, labeled "Online Learning Workflow with LearnMate²," shows the three corresponding solutions: (A) PlanMate generates a personalized learning schedule based on goals, time, path, and pace; (B) StudyMate provides real-time contextual assistance alongside course video; and (C) AdaptMate generates an adapted learning schedule based on quiz performance.}
   \vspace{-12pt}
   \caption{\textit{From fixed online learning workflows to personalized and adaptive support ---} The figure contrasts a typical online learning workflow (left), where learners often struggle with cold start problems (\protect\darkgreysquare{A}), limited timely support (\protect\darkgreysquare{B}), and fixed learning plans (\protect\darkgreysquare{C}), with the \ours{} workflow (right). \ours{} addresses these challenges through \texttt{PlanMate} for personalized study planning (\protect\darkgreysquare{A}), \texttt{StudyMate} for real-time contextual assistance (\protect\darkgreysquare{B}), and \texttt{AdaptMate} for adaptive learning activities based on quiz performance and progress (\protect\darkgreysquare{C}).}
  \label{fig:system_workflow}
%   \vspace{-6pt}
\end{teaserfigure}

\ccsdesc[500]{Human-centered computing~HCI design and evaluation methods}
\ccsdesc[300]{Computing methodologies~Natural language processing}

%%
%% Keywords. The author(s) should pick words that accurately describe
%% the work being presented. Separate the keywords with commas.
\keywords{large-language models; human-in-the-loop; human-centered AI}

%\received{20 February 2007}
%\received[revised]{12 March 2009}
%\received[accepted]{5 June 2009}

%%
%% This command processes the author and affiliation and title
%% information and builds the first part of the formatted document.

% \begin{teaserfigure}
% %     \includegraphics[width=\textwidth]{figures/teaser new.pdf}
% %    \vspace{-12pt}
% %   \caption{\textit{Outline of the MAP framework and \ours{} ---} In this paper, we draw from lifelong learning theories of the education domain to design a framework that supports AI systems in providing personalization to multiple users across three stages: \textit{reflection}, \textit{articulation}, and \textit{feedback}. We implement our framework using multiple LLM agents that operate within a hierarchical structure, employing retrieval-augmented generation (RAG) techniques.
% % }
%   \label{fig:teaser}
% \end{teaserfigure}
%\begin{teaserfigure}
%    % \includegraphics[width=\textwidth]{figures/VERL_teaser copy copy.pdf}
%    % \vspace{-12pt}
%    % \caption{\textit{\ours{} --- } xxx}
%    % \label{fig:teasor}
%\end{teaserfigure}
\maketitle

\section{Introduction}
%P1: Introduce online learning and problems in online learning

Online education has transformed access to learning by making educational content available across geographic, economic, and scheduling constraints \cite{kumar2019online, money2019incorporating, iniesto2024understanding, badge2008assessing, veletsianos2019analysis, black2019online}. Online platforms now support millions of learners, from students in underserved areas to working professionals pursuing continuing education, through resources such as video lectures, tutorials, digital textbooks, and curated articles \cite{pardamean2022ai}. As a result, online learning has become an essential part of modern education.

%P2: lack of personalzation in online learning

Despite these benefits, current online learning platforms often provide limited personalized support. Many still adopt a one-size-fits-all model, presenting standardized content without adequately accounting for differences in learners' goals, pace, progress, or support needs \cite{gillett2017challenges}. This limitation is reflected in persistent challenges such as low completion rates and reduced satisfaction in large-scale online courses, where learners often receive little individualized guidance \cite{daniel2012making}. Prior work points to three recurring gaps in online learning environments. First, learners often struggle to plan and regulate their study process without instructor guidance, leading to difficulties with time management and goal setting \cite{zimmerman2002becoming, bernacki2021systematic}. Second, learners often lack timely instructional support while actively studying, making it difficult to receive clarification or scaffolding when confusion arises \cite{taber2018scaffolding, SPOUSE1998259, hung2022scaffolding}. Third, many systems collect performance data but do not effectively translate it into adaptive interventions that respond to learners' ongoing progress \cite{maier2022personalized, murtaza2022ai}.

% P4: AI's solution and problems

Recent advances in artificial intelligence (AI), particularly large language models (LLMs), create new opportunities to address these gaps. Emerging AI-based educational systems have shown promise in areas such as adaptive feedback, personalized content generation, and in-situ learning support \cite{10.1145/3613904.3642773, 10.1145/3706598.3713275, 10.1145/3706599.3720240}. However, it remains unclear how such capabilities should be integrated into online learning workflows as a coordinated form of personalization rather than as isolated features. In particular, the effects of different LLM-supported personalization mechanisms on learner performance and user experience remain underexplored. Recent work has begun to explore LLM-based personalization in online learning \cite{wang2025learnmate, park2024empowering, wen2024ai},
% {\color{blue}{\cite{wang2025learnmate, park2024empowering, wen2024ai}}},
providing a foundation for studying how these mechanisms can be designed and evaluated more systematically.

%P5: Our work 

In this paper, we extend the personalization framework of \citet{wang2025learnmate} through \tool{}, an LLM-powered personalized and adaptive learning support system for online learning. \tool{} integrates three components that correspond to the gaps above: (1) \texttt{PlanMate}, which generates personalized study plans based on four dimensions---goals, time, pace, and path; (2) \texttt{StudyMate}, which provides real-time contextual assistance during learning sessions through progressive disclosure, follow-up explanations, practice questions, and external resources; and (3) \texttt{AdaptMate}, which adjusts learning activities and study plans based on quiz performance and learning progress. Together, these components form a closed-loop learning workflow in which planning, studying, and adaptation continuously inform one another. 
% {\color{blue}{At a high level, we investigate, \textit{what are the overall effects of personalization on learning outcomes in online learning?} More specifically, this paper addresses the following research questions:}}
At a high level, we investigate, \textit{what are the overall effects of personalization on learning outcomes in online learning?} More specifically, this paper addresses the following research questions:

% \begin{enumerate}
%     \item {\color{blue}{$RQ_1$: How do \textit{personalized study plans} support learners' planning and learning experience, specifically usability and satisfaction?}}
%     \item {\color{blue}{$RQ_2$: How does \textit{real-time contextual assistance} during learning sessions support learners' performance and learning experience, particularly learners' quiz outcomes, usability and satisfaction?}}
%     \item {\color{blue}{$RQ_3$: How do \textit{adaptive learning activities} affect learners' progress and learning experience, notably usability and satisfaction?}}
%     \item {\color{blue}{$RQ_4$: How does a personalization workflow that integrates these components affect these learning outcomes and experience?}}
% \end{enumerate}

% \begin{enumerate}
%     \item {\color{blue}{\textit{RQ1}: How do personalized study plans support learners' planning and learning experience?}}
%     \item {\color{blue}{\textit{RQ2}: How does real-time contextual assistance during learning sessions support learners' performance and learning experience?}}
%     \item {\color{blue}{\textit{RQ3}: How do adaptive learning activities affect learners' progress and learning experience?}}
%     \item {\color{blue}{\textit{RQ4}: How does a personalization workflow that integrates these components affect the learner's learning outcomes and experience?}}
% \end{enumerate}

\begin{enumerate}
    \item \textit{RQ1}: How do personalized study plans support learners' planning and learning experience?
    \item \textit{RQ2}: How does real-time contextual assistance during learning sessions support learners' performance and learning experience?
    \item \textit{RQ3}: How do adaptive learning activities affect learners' progress and learning experience?
    \item \textit{RQ4}: How does a personalization workflow that integrates these components affect the learner's learning outcomes and experience?
\end{enumerate}

To answer these research questions, we designed and developed \tool{} and conducted two studies. 
% {\color{blue}{Across these studies, we examine learning performance primarily through quiz performance and learning experience through participants' reported usability and satisfaction. }}
Across these studies, we examine learning performance primarily through quiz performance and learning experience through participants' reported usability and satisfaction.
We first conducted a preliminary study ($n=24$) to assess how well the system supported personalized learning and to identify usability issues. We then refined the system and conducted a final user study ($n=16$) comparing \tool{} against a workflow combining a state-of-the-art online learning platform and an LLM used for learning support. Our results show that \tool{} improves quiz performance and user experience compared to existing online learning and support tools. Our work makes the following contributions:

\begin{enumerate}
    \item \textit{System contribution:} We present \tool{}, an LLM-based learning system that integrates personalized study planning based on four-dimensional guidelines (\texttt{PlanMate}), course-grounded real-time contextual assistance with progressive disclosure (\texttt{StudyMate}), and formative plan adaptation based on quiz performance and learning progress (\texttt{AdaptMate}).
    \item \textit{Empirical contribution:} We evaluate \tool{} through a preliminary study ($n=24$) and a user study ($n=16$) to understand its effectiveness, the contributions of its individual components, and the value of the integrated workflow.
    \item \textit{Design implications:} We present design implications for AI-driven educational tools that conceptualize personalization as a continuous learning workflow, emphasize course-grounded and in-situ support, and enable mixed-initiative, learner-steerable adaptation. 
\end{enumerate}

\section{Related Works}

\subsection{Personalized Learning in Online Environments}
Personalized learning is defined as ``instruction that is paced to learning needs, tailored to learning preferences, and tailored to the specific interests of different learners'' \cite{usdoe2010}. For digital learning contexts specifically, \citet{bernacki2021systematic} characterizes personalized learning environments as systems that must be: (1) adaptive to individual knowledge, experience, and interests, and (2) effective and efficient in supporting and promoting desired learning outcomes.
In 2022, \citet{maier2022personalized} developed an analytic framework for personalized feedback implementations that distinguishes three hierarchical levels: microscale feedback (referring to one item or assignment), mesoscale feedback (referring to specific learning goals or assessments), and macroscale feedback (referring to extended learning periods such as weekly feedback during a semester).

Personalized learning offers several crucial benefits, including improved learning effectiveness through enhanced interactive engagement and real-time adaptive content delivery \cite{swan2003learning, spaho2025iot}, increased accountability through transparent learning analytics and explainable recommendation mechanisms that enable learners to track their progress and understand the rationale behind personalized content delivery \cite{elwarraki2023toward}, broader accessibility through flexible learning pathways and device-agnostic platforms that accommodate diverse learner needs, socioeconomic contexts, and technological constraints \cite{al2024adaptive, murtaza2022ai}, 
% and enhanced efficiency through personalized recommendation systems that automatically identify and deliver relevant learning resources while eliminating manual content discovery \cite{bin2024comprehensive}.
and enhanced efficiency through personalized recommendation systems that identify and deliver relevant learning resources, eliminating manual content discovery.

% Current Approaches and Limitations & Challenges
% {\color{blue}{Prior to LLM-based approaches, personalization in online learning was primarily achieved through structured adaptive systems and rule-based learning technologies. At the learner-task level, systems such as Cognitive Tutors provided step-by-step instructional guidance tailored to individual performance, demonstrating how adaptation could support learners during problem solving \cite{koedinger2007exploring, aleven2016instruction}. At the course level, frameworks such as MOOClet \cite{reza2021mooclet} enabled instructors and platform designers to personalize aspects of online courses through experimentation and performance-based adjustments.}}
Prior to LLM-based approaches, personalization in online learning was primarily achieved through structured adaptive systems and rule-based learning technologies. At the learner-task level, systems such as Cognitive Tutors provided step-by-step instructional guidance tailored to individual performance, demonstrating how adaptation could support learners during problem solving \cite{koedinger2007exploring, aleven2016instruction}. At the course level, frameworks such as MOOClet \cite{reza2021mooclet} enabled instructors and platform designers to personalize aspects of online courses through experimentation and performance-based adjustments.
More broadly, online learning platforms have incorporated foundational personalization features such as content recommendation based on learning history, progress tracking, and rule-based feedback \cite{bernacki2021systematic, shemshack2020systematic}. 
For example, Coursera \cite{coursera2026} offers course recommendations and limited quiz-based branching, while OnTask \cite{ontask2026} supports instructor-configured personalized feedback based on assessment and engagement data \cite{murtaza2022ai, rader2018explanations}. 

These systems show that personalization in online learning is not new. However, prior approaches have largely depended on pre-authored rules, historical usage patterns, or instructor-configured interventions, which limits their ability to provide flexible, real-time, and context-sensitive support across the full learning workflow. As a result, two important gaps remain: learners often face a cold start problem and lack structured guidance when beginning a new subject, and many systems provide static, rule-based support rather than real-time assistance during learning sessions \cite{maier2022personalized, murtaza2022ai, kem2022personalised, 10445145}.

%In this work, we address these critical limitations through an integrated approach. To solve the cold start problem and lack of structured guidance, we provide systematic four-dimensional personalization guidance (\textit{goals}, \textit{time}, \textit{pace}, \textit{path}) that helps learners create concrete study plans even without prior domain knowledge, thereby improving efficiency and accountability in the planning process. To address static, rule-based systems without real-time support, we offer dynamic contextual assistance during active learning sessions that adapts to learners' immediate needs and questions, enhancing both effectiveness and accessibility of online learning. 

% effectiveness
% accessibliy
% efficieny
% accountabily
\subsection{Adaptive Learning Systems and Activities}

Adaptive learning is an educational technology that dynamically adjusts instructional content to provide personalized learning paths for individual learners \cite{martin2020systematic}. 
This approach contributes to online education in two primary ways: (1) enhances learner confidence through calibrated challenges and personalized feedback that promote positive self-judgments of competence and capability \cite{maclellan2014might}. (2) improves learning autonomy through self-organizing systems that enable learners to independently discover critical features, conduct hypothesis testing, and adapt without external supervision \cite{grossberg2020path, iida2023timescales}.

Traditional adaptive learning platforms often follow rule-based approaches, presenting predetermined content sequences based on simple performance metrics without considering learners' unique characteristics and preferences \cite{gligorea2023adaptive}. 
These systems face significant challenges, including cognitive bottlenecks, disciplinary fragmentation, and limitations in emotional modeling that prevent truly intelligent learning experiences \cite{li2021progress}. Modern AI-driven adaptive learning systems use data-driven algorithms, machine learning, and big data analytics to dynamically adjust content, delivery, and instructional pace based on individual learner characteristics, performance data, and behavioral patterns \cite{gligorea2023adaptive, kem2022personalised, minn2022ai}. For example, systems now employ deep knowledge tracing to model student knowledge states and predict performance, while platforms like Carnegie Learning utilize AI-powered technology to generate personalized assignments \cite{minn2022ai, vandewaetere2011contribution}. 

However, despite rapid development in AI adaptive learning technologies, current implementations face significant limitations. First, there remains a lack of comprehensive research on actual deployment and effectiveness evaluation in online learning environments, with most work confined to theoretical frameworks rather than empirical validation \cite{vandewaetere2011contribution, gligorea2023adaptive}. Second, most systems continue to depend on instructor input rather than providing fully autonomous learning support that can adapt based on ongoing performance assessment and learning progress \cite{imran2024personalization, maier2022personalized}.
%In this work, to overcome limited autonomous adaptation, we implement assessment-driven plan modifications that function independently of instructor input, creating a complete learning loop that evolves with learner progress while maintaining transparency and learner control, thereby improving confidence and autonomy.

\subsection{AI and LLM Applications in Education}

AI has been extensively adopted in education across three key areas: administration (automated grading and feedback), instruction (intelligent tutoring systems), and learning (personalized content delivery), offering potential solutions to personalization challenges in online learning \cite{chen2020artificial}. These applications could potentially provide improved efficiency in administrative tasks, enhanced instructional quality through customization, and personalized learning experiences that foster better information retention.

Recent LLM research demonstrates diverse educational applications across multiple domains. Instructional support systems provide step-by-step programming guidance, conversational AI for collaborative learning, and interactive learning environments for children \cite{10.1145/3613904.3642773, 10.1145/3706599.3720240, 10.1145/3613905.3650868, 10.1145/3613904.3642229}. 
Content personalization applications include adaptive story reading experiences, AI-generated educational podcasts, specialized learning support for diverse populations, and project-based learning assistance \cite{10.1145/3706598.3713275, 10.1145/3706598.3713460, 10.1145/3613904.3642899, 10.1145/3706598.3713971}. 
Adaptive learning systems incorporate human-LLM collaboration, biosignal-driven learning companions, and assessment-driven content adjustment \cite{10.1145/3706599.3719877, 10.1145/3706599.3719852, 10.1145/3613904.3642366}. Advanced system architectures employ multi-agent approaches and Retrieval-Augmented Generation techniques to maintain coherent, context-aware interactions \cite{qin2023toolllm, lewis2020retrieval}.
% {\color{blue}{Relevant to our work, \citet{fan2024lessonplanner} demonstrated how LLMs can scaffold structured planning tasks, while \citet{leong2024putting} showed that context-grounded personalization improves learner motivation.}}
Relevant to our work, \citet{fan2024lessonplanner} demonstrated how LLMs can scaffold structured planning tasks, while \citet{leong2024putting} showed that context-grounded personalization improves learner motivation.

However, significant gaps remain: most existing systems provide support for individual educational tasks, with limited exploration of how these task-level supports can be integrated to scaffold learners across an entire learning workflow, particularly in online learning contexts.
While \citet{wang2025learnmate} explored personalization frameworks using LLMs, their work focused on system design without empirical evaluation of learning effectiveness. This represents a critical opportunity for advancing the field through comprehensive evaluation of integrated systems that combine personalized planning, real-time assistance, and adaptive activities within unified learning environments.

In summary, our work aims to support and improve online learning outcomes and user experience, specifically efficiency, accountability, accessibility, and effectiveness through three integrated components: personalized study planning (\texttt{PlanMate}) that provides structured four-dimensional guidance to address cold start problems, real-time contextual assistance (\texttt{StudyMate}) that delivers dynamic support during active learning sessions, and adaptive learning activities (\texttt{AdaptMate}) that enable assessment-driven plan modifications to create autonomous learning loops.

\section{Our System: \tool{}} 
\label{sec:system}

Building on prior work by \citet{wang2025learnmate}, we developed \tool{}, an LLM-based system for personalized and adaptive online learning. \tool{} comprises three interconnected components: \texttt{PlanMate}, which supports personalized study planning; \texttt{StudyMate}, which provides real-time contextual assistance during learning sessions; and \texttt{AdaptMate}, which adapts study plans based on learner progress and quiz performance. \texttt{PlanMate} and \texttt{StudyMate} extend the prior \texttt{LearnMate} framework, while \texttt{AdaptMate} was introduced following insights from our preliminary study.

This section describes \tool{}'s system architecture, user interaction workflow, and technical implementation. We first present a guiding user scenario, then describe the integrated learning loop, and finally detail how personalization is implemented across the three components.

\begin{figure*}[!tb]
  \includegraphics[width=\textwidth]{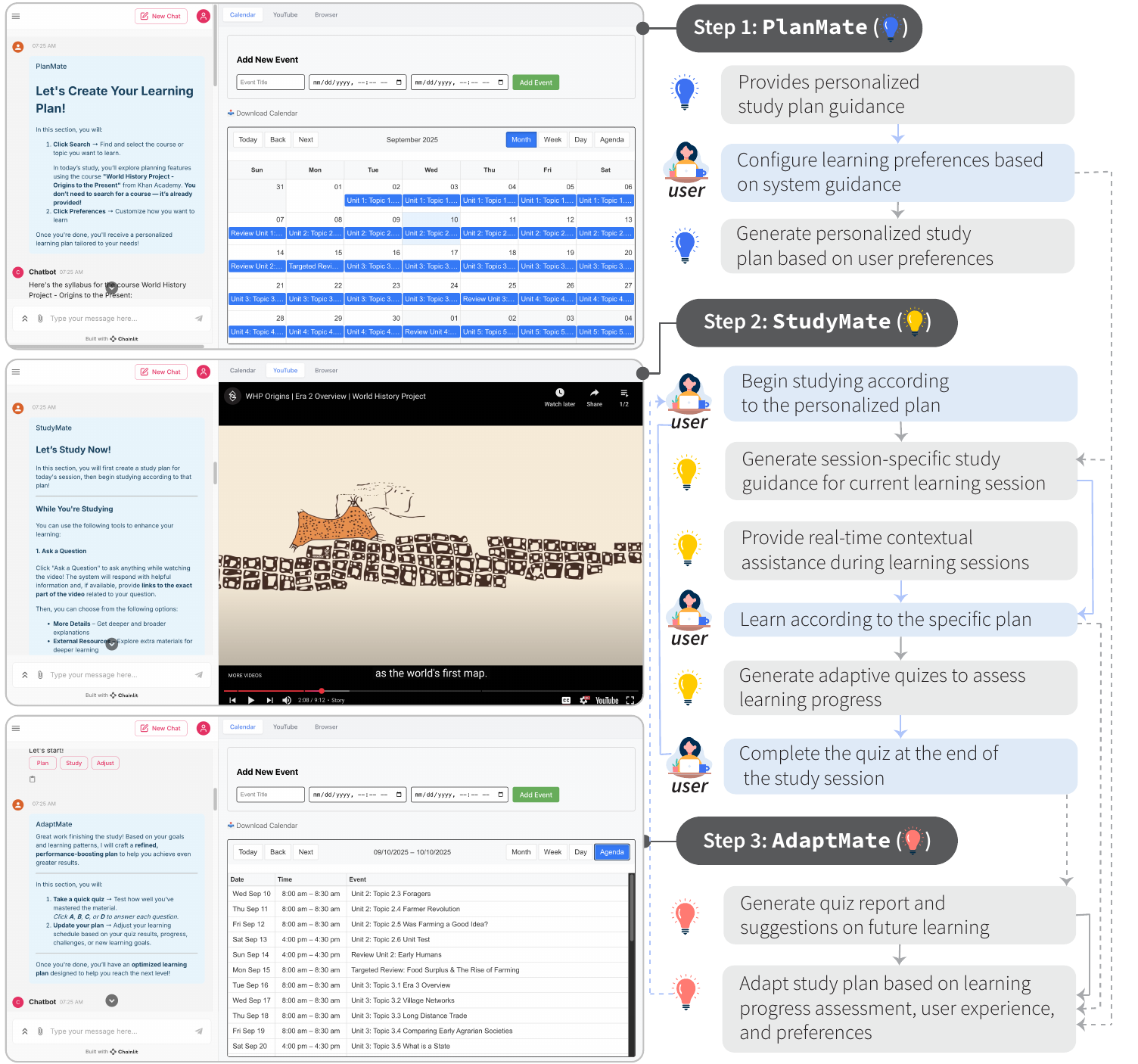}
   \vspace{-12pt}
  \caption{\textit{\ours{} System Workflow ---} 
  The front-end interface of \tool{} (left) and the system workflow (right). The three-step integrated learning loop shows how \texttt{PlanMate}, \texttt{StudyMate}, and \texttt{AdaptMate} work together to provide personalized study planning, real-time contextual assistance, and adaptive learning activities that adjust based on user progress and performance.
  We outline and explain the pipeline of \tool{} in Section~\ref{sec:system}.
  \textit{Screenshot © Khan Academy (khanacademy.org), used under CC BY-NC-SA 4.0.}}
  \label{fig:system_workflow}
  \Description[LearnMate² System Workflow]{A diagram showing the LearnMate² front-end interface on the left and the system pipeline on the right. The pipeline is divided into three steps. Step 1, PlanMate, shows the system providing personalized study plan guidance, configuring learning preferences based on user input, and generating a personalized study plan displayed in a calendar. Step 2, StudyMate, shows the user beginning a study session, receiving session-specific guidance, getting real-time contextual assistance, completing learning, and taking an adaptive quiz. Step 3, AdaptMate, shows the system generating a quiz report and adapting the study plan based on performance, experience, and preferences.}
\end{figure*}

\subsection{Guiding User Scenario}
To illustrate \tool{}'s functionality, we present a scenario featuring Alex, a high school student from a rural community with limited access to advanced coursework who is preparing for the AP World History exam through self-directed online learning.

\begin{quote}
    \textit{Alex is a junior at a small rural high school that doesn't offer AP World History due to limited resources and staffing. Determined to take the AP exam to strengthen college applications, Alex must rely entirely on online resources for preparation. With approximately 6-8 hours per week available for study (balancing school, part-time work, and family responsibilities), Alex needs an efficient and structured approach to cover the extensive AP curriculum. Having primarily learned through traditional textbook-based methods, Alex is unfamiliar with optimizing online learning strategies and seeks guidance on creating an effective study plan that can adapt to varying schedules and learning progress.}
\end{quote}

% This scenario demonstrates how \tool{} can assist learner's oline learning useing the threee key features. 
This scenario demonstrates how \tool{} assists learners in online learning through the three key features.
% learning experience and learning bridge educational equity gaps by providing personalized learning support to students who lack access to traditional AP preparation resources, illustrating the system's potential for democratizing quality education.

% This scenario demonstrates how \tool{} can bridge educational equity gaps by providing personalized learning support to students who lack access to traditional AP preparation resources, illustrating the system's potential for democratizing quality education.

% \begin{quote}
%     \textit{astronmy?/or the world history project as in user study}

% \end{quote}

\subsection{System Overview: Integrated Learning Loop}
% \tool{} implements a closed-loop personalized and adaptive learning architecture comprising three interconnected components that work together to create a dynamic, personalized learning experience. Unlike traditional linear approaches, our system establishes a continuous feedback cycle: \texttt{PlanMate} generates personalized study plans, \texttt{StudyMate} provides real-time learning support and assessment, and \texttt{AdaptMate} adjusts plans based on performance data, creating an iterative learning loop that evolves with the learner's progress.

% \tool{} implements a closed-loop personalized and adaptive learning architecture comprising three interconnected components: (1)\texttt{PlanMate}, (2)\texttt{StudyMate}, and (3)\texttt{AdaptMate} .

\tool{} implements a closed-loop personalized and adaptive learning architecture comprising three interconnected components: (1) \texttt{PlanMate}, (2) \texttt{StudyMate}, and (3) \texttt{AdaptMate}.

\paragraph{\texttt{PlanMate}}

\texttt{PlanMate} creates customized learning plans based on the four-dimensional personalization framework (\textit{goals}, \textit{time}, \textit{pace}, and \textit{path}) established in existing literature on personalization in online learning \cite{graham2019k} and recent work on AI-based personalization \cite{wang2025learnmate}. The component analyzes user preferences and course requirements to generate structured study schedules with appropriate content sequencing and time allocation. 
\texttt{PlanMate} also accepts adaptation signals from \texttt{AdaptMate} to regenerate optimized plans based on learning outcomes.

\paragraph{\texttt{StudyMate}}
% \texttt{StudyMate} provides comprehensive learning assistance during study sessions through three key mechanisms: (1) contextual question-answering that draws from course materials and learning history, (2) progressive disclosure framework offering more details, practice questions , and external resoucrse  (3) formative assessment through adaptive quizzes that evaluate comprehension and identify knowledge gaps. 
% \texttt{StudyMate} continuously monitors learning interactions to gather performance data for the adaptation cycle.

\begin{figure*}[!th]
  \includegraphics[width=\textwidth]{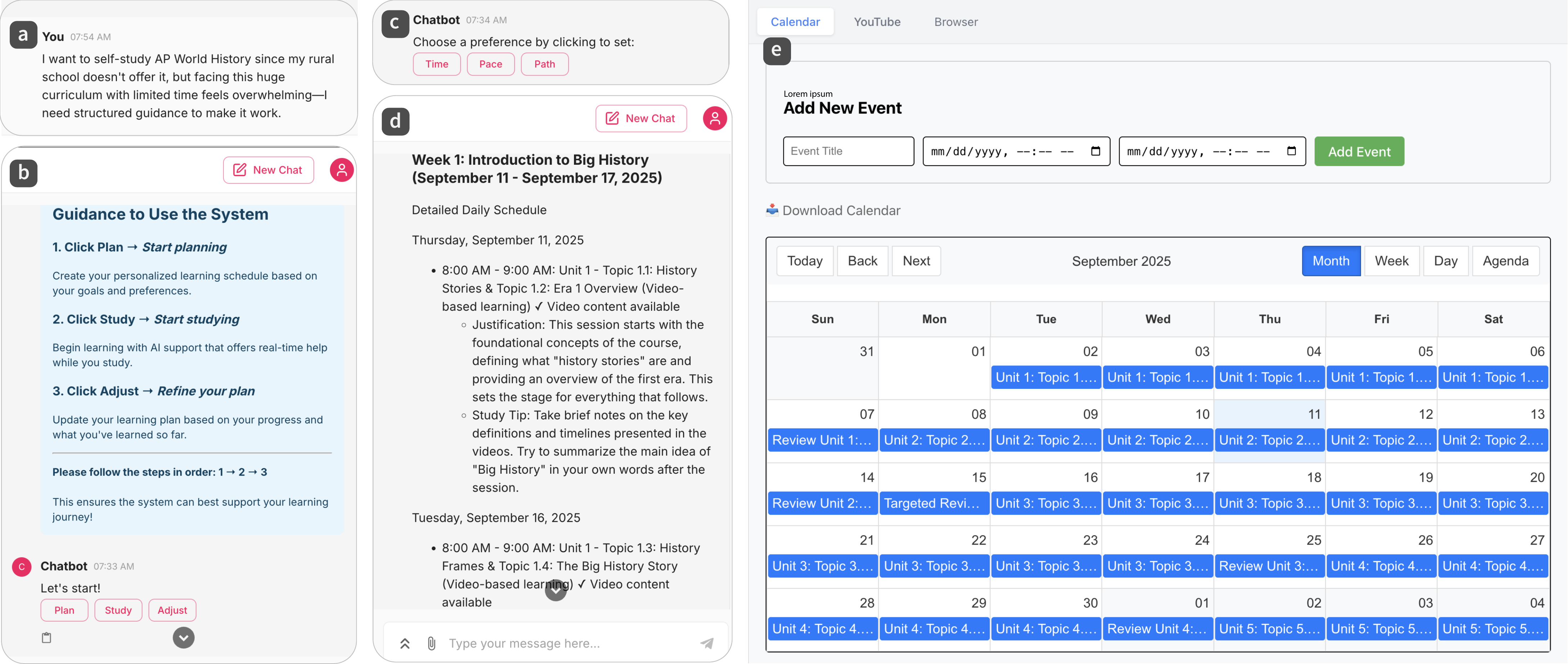}
   \vspace{-12pt}
  \caption{\textit{\texttt{PlanMate} Interface ---} The front-end interface for personalized study planning. Users begin by specifying their learning goals in natural language (step \protect\darkgreysquare{a}), then provide preferences across multiple dimensions, including time availability, learning pace, and content path (steps \protect\darkgreysquare{b} to \protect\darkgreysquare{c}). Based on these inputs, \texttt{PlanMate} generates a personalized study plan in text form (step \protect\darkgreysquare{d}) and visualizes it as an interactive calendar schedule (step \protect\darkgreysquare{e}).We outline the user's interaction with the front-end as a guide to explain the pipeline of \texttt{PlanMate} in Section \ref{sec:planmate}.}
  \label{fig:plan_system}
  \Description[PlanMate Interface]{A screenshot of the PlanMate interface with five labeled regions. (a) A chat panel where a user types their learning goal in natural language, stating they want to self-study AP World History but feel overwhelmed by the curriculum. (b) A system guidance panel listing three steps: Click Plan to start planning, Click Study to start studying, and Click Adjust to refine the plan. (c) A chatbot response offering three preference buttons: Time, Pace, and Path. (d) A generated weekly study plan for "Week 1: Introduction to Big History" with a detailed daily schedule including specific topics and study tips. (e) An interactive calendar view showing September 2025 with color-coded study sessions populated across the month, including units, topics, and review sessions.}
\end{figure*}

\texttt{StudyMate} provides comprehensive learning assistance during study sessions through three key mechanisms: (1) contextual question-answering that draws from course materials and learning history; (2) progressive disclosure framework that offers more details, practice questions, and external resources; and (3) formative assessment through adaptive quizzes that evaluate comprehension and identify knowledge gaps. 
\texttt{StudyMate} continuously monitors learning interactions to gather performance data for the adaptation cycle.

% 1.1 more details
% 1.2 practice questions 
% 1.3 external resoucrse

\paragraph{\texttt{AdaptMate}}
Introduced following insights from the preliminary study, \texttt{AdaptMate} analyzes learning performance data and interactions from \texttt{StudyMate} to identify areas requiring adjustment, then generates modifications to the personalized plan based on this analysis.
The component evaluates quiz results, interaction patterns with the system during learning sessions, and four-dimensional learner preferences to determine optimal modifications such as additional time allocation for challenging topics, alternative learning resources, or adjusted pacing. 
\texttt{AdaptMate} then communicates these insights back to \texttt{PlanMate} to regenerate improved learning plans, completing the adaptive loop.

% \paragraph{Integrated Learning Cycle}
% The three components operate as an integrated system where each phase informs and improves the others. This creates a responsive learning environment that adapts to individual progress patterns, maintains engagement through appropriate challenge levels, and optimizes learning outcomes through continuous refinement of the educational experience.

%\paragraph{Contextual Q\&A System}

% \paragraph{Progressive Disclosure Framework}  
% 1.1 more details
% 1.2 practice questions 
% 1.3 external resoucrse

% \paragraph{Formative Assessment Integration}
% 2. quiz to access overall understanding
% 2.1 quiz
% 2.2 summary and suggestion of quiz data

% \subsection{User Interaction with \tool{} and workflow}
% in this theme, we will each detailed introduce user interaction with three main features, followedby how it works/implementation as shown in \ref{fig:system_workflow}.

% \subsubsection{\texttt{PlanMate}: Enhanced Personalized Planning}
% alex will first chooes the course and tell the system on the right pannel by typing (goals). then the system provide guidance, alex will input his prefernces by clicking ecach button, \textit{time}, \textit{pace}, \textit{path}, then input. then system generates the plan

% 1. input guideline by \cite{} (goals, time, pace, path)
% 2. genrate

\subsection{User Interaction with \tool{}}
\subsubsection{Personalization in \tool{}}

Before detailing each component, we clarify how personalization is implemented across \tool{}'s learning workflow. \tool{} supports three distinct but connected forms of personalization, each operating at a different stage of learning.

\paragraph{\texttt{PlanMate}: preference-based planning personalization.} \texttt{PlanMate} personalizes the study schedule along four dimensions adapted from prior work~\cite{graham2019k}: learning goals, time availability, pace, and content path. Users explicitly provide these preferences through dedicated interface controls and natural-language inputs (\eg ``I have 2 hours on weekdays and prefer to go slowly''), which the system uses to generate an individualized study plan.

\paragraph{\texttt{StudyMate}: context- and history-aware in-session personalization.} \texttt{StudyMate} personalizes support during active study sessions using the learner's accumulated history, including prior session content, quiz performance, and the goals and preferences established in \texttt{PlanMate}. Users further shape this support in real time by asking questions during the session, allowing the system to tailor responses to the current material and the learner's immediate needs.

\paragraph{\texttt{AdaptMate}: performance-driven adaptive personalization.} \texttt{AdaptMate} adapts the study plan based on the learner's quiz performance, interaction patterns during study (\eg help-seeking frequency and difficult topics), content coverage, and the four-dimensional preferences previously provided in \texttt{PlanMate}. Users initiate this adaptation by clicking \textit{Generate Adjusted Plan}, after which the system proposes targeted modifications such as reallocating time, reordering content, or recommending alternative resources.

\subsubsection{\texttt{PlanMate}: Enhanced Personalized Planning}
\label{sec:planmate}

The front-end interface of \texttt{PlanMate} is shown in Figure~\ref{fig:plan_system}. Alex begins by typing \textit{``I want to learn the World History Project course from Khan Academy''} to define their learning goals through the interface on the right panel (depicted as step \darkgreysquare{a}. 
The system then provides guidance prompts where Alex inputs their preferences by clicking each dimension button: \textit{time availability}, \textit{learning pace}, and \textit{content path preferences}. Alex can specify their preferences in natural language for each dimension by clicking the corresponding button (step \darkgreysquare{b}, \darkgreysquare{c}).
Once all the preferences are inputted, the system automatically creates a personalized study schedule in text format (step \darkgreysquare{d}) and displays it in an interactive calendar interface (\darkgreysquare{e}).

\paragraph{Technical Details. }

\texttt{PlanMate} utilizes a two-stage generation process, each with a separate agent powered by \texttt{Gemini-2.0-Pro} using carefully crafted prompts designed for the specific task. 
The first stage analyzes user inputs across the four personalization dimensions (\textit{goals}, \textit{time}, \textit{pace}, \textit{path}) using structured prompts that incorporate educational pedagogy principles from established literature \cite{graham2019k}.
% The second stage converts the generated plan into calendar events using \texttt{ICS} format and displays in React Big Calendar \cite{reactbigcalendar}, enabling interactive schedule management through drag-and-drop functionality.
The second stage converts the plan into \texttt{ICS} calendar events displayed in React Big Calendar \cite{reactbigcalendar}, supporting drag-and-drop schedule management.

\subsubsection{\texttt{StudyMate}: Interactive Learning Support}
\label{studymate}

\begin{figure*}[!th]
  \includegraphics[width=\textwidth]{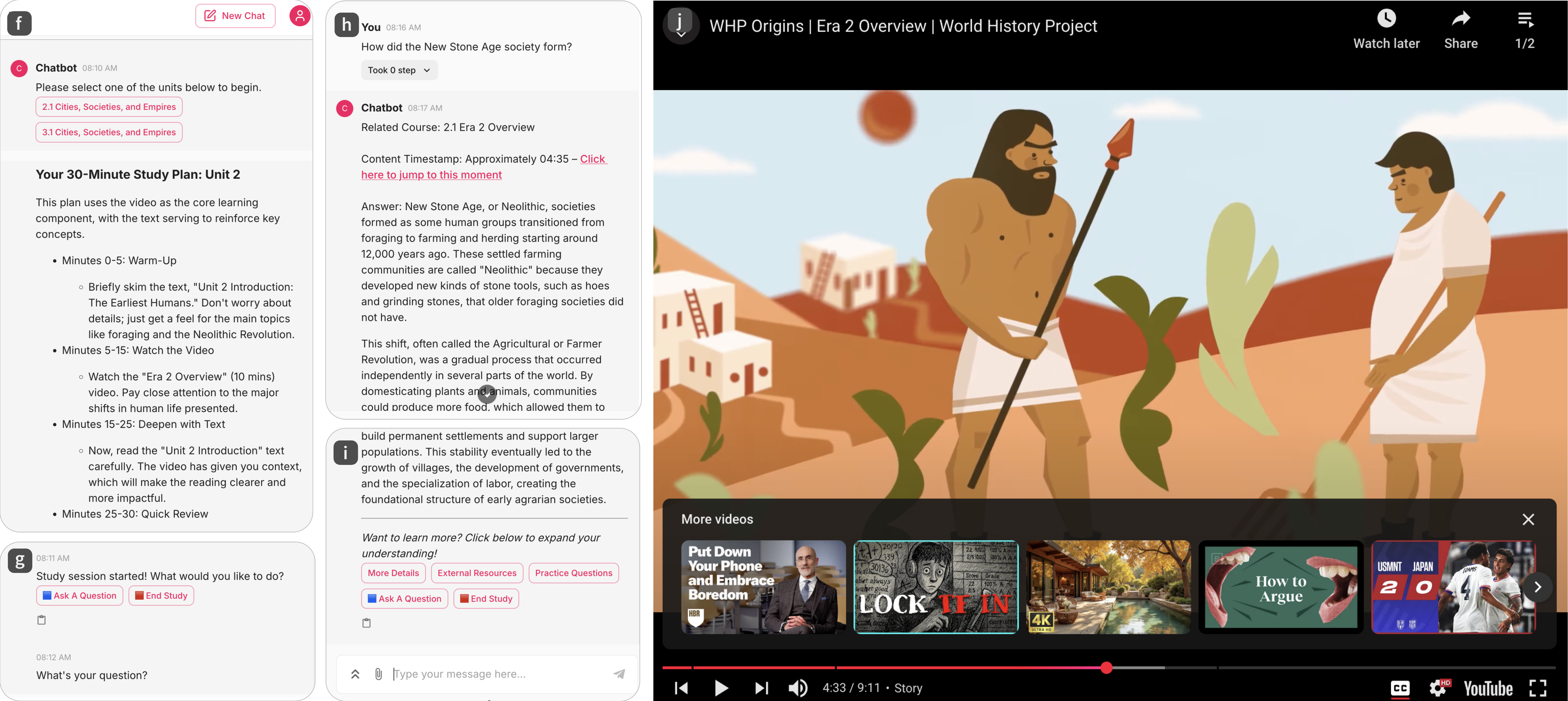}
   \vspace{-12pt}
  \caption{\textit{\texttt{StudyMate} Interface ---} The front-end interface for interactive learning support during scheduled study sessions. Users start a learning session by selecting a planned time slot (step \protect\darkgreysquare{f}), after which \texttt{StudyMate} provides session-specific guidance. During the session, users can ask questions and receive contextual assistance with progressive disclosure options, including more details, practice questions, and external resources (steps \protect\darkgreysquare{g} to \protect\darkgreysquare{i}). At the end of the session, \texttt{StudyMate} generates an adaptive quiz to assess learning outcomes (step \protect\darkgreysquare{k}). We outline the user's interaction with the front-end as a guide to explain the pipeline of \texttt{StudyMate} in Section \ref{studymate}.
  \textit{Screenshot © Khan Academy (khanacademy.org), used under CC BY-NC-SA 4.0.}}
  \label{fig:study_system}
  \Description[StudyMate Interface]{A split-screen screenshot. The left panel shows a chat interface. (f) The chatbot prompts the user to select a unit to begin, offering "2.1 Cities, Societies, and Empires" and "3.1 Cities, Societies, and Empires," and displays a 30-minute study plan. (g) The study session starts with buttons for "Ask a Question" and "End Study." (h) A user asks "How did the New Stone Age society form?" (i) The chatbot responds with a contextually grounded answer citing course content at timestamp 4:35, explaining the Neolithic transition, with expandable options for More Details, External Resources, and Practice Questions. The right panel (j) shows a YouTube video player displaying a Khan Academy World History Project video "WHP Origins | Era 2 Overview" paused at 4:33, with an illustration of early human figures.}
\end{figure*}

The front-end interface of \texttt{StudyMate} is shown in Figure~\ref{fig:study_system}. Alex initiates their first study session by clicking the scheduled session button (\eg \textit{Sep 7, 8pm-9pm: Unit 2.1 Cities, Societies, and Empires}) from their generated plan. \texttt{StudyMate} immediately generates and provides session-specific study guidance based on the selected time slot, learning objectives, and preferences established in \texttt{PlanMate} (step \darkgreysquare{f}). 
Alex then initiates the learning session through the right panel interface (step \darkgreysquare{j}). 
During learning, Alex can ask questions by clicking \textit{``Ask a Question''} (step \darkgreysquare{g}) and typing the question to receive contextual assistance (step \darkgreysquare{h}). Each response utilizes a progressive disclosure framework offering three expandable options: \textit{``more details''}, \textit{``practice questions''}, and \textit{``external resources''} (step \darkgreysquare{i}). 
Upon completing the study session, Alex clicks \textit{``End Study''} which triggers an adaptive quiz with four-answer multiple-choice questions based on the covered content (step \darkgreysquare{k}). 
Alex can choose to relearn the session if needed or continue studying following the same pattern for subsequent sessions.
% Alex can choose to relearn the session if needed or continue studying following the same pattern for subsequent sessions by clicking the ssesion button?.

\paragraph{Technical Details. } 

\texttt{StudyMate} integrates content analysis and user modeling to provide contextual support through three specialized agents: a question-answering agent, a quiz generation agent, and a quiz analysis agent. The system processes course transcripts and maintains a learning history database to generate relevant responses.
The question-answering agent uses hierarchical prompting strategies where each tier builds upon the previous response while incorporating different pedagogical approaches (more detailed explanation, practice questions, and extension material). When users ask questions, the agent first provides a concise answer based on the course content, then offers expandable options for deeper exploration.
% {\color{blue}{
% When users submit off-topic or out-of-scope requests, the system still provides a response but explicitly notes that the question falls outside the current session's learning content.
% }}
When users submit off-topic or out-of-scope requests, the system still provides a response but explicitly notes that the question falls outside the current session's learning content.
% The quiz generation agent employs content analysis algorithms combined with example questions from existing learning materials to identify key concepts from the study session. This agent creates four multiple-choice questions that assess comprehension of the most important topics covered during the learning period.
The quiz generation agent employs structured prompting techniques combined with example questions from existing learning materials to identify key concepts from the study session. This agent creates four multiple-choice questions that assess comprehension of the most important topics covered during the learning period.
The assessment agent analyzes user interaction data, quiz outcomes, and learned content to evaluate responses against learning objectives, then sends the results to the adaptive agent in \texttt{AdaptMate}.
% the assessment agaent useuser intection data, quiz outcome, and learnt contetn to evaluates responses against learning objectives to provide targeted feedback and recommendations.

% \subsubsection{\texttt{AdaptMate}: Intelligent Plan Adaptation} 
% alesnow can click the button ``genearted adjusted plan'' beacuse he feels the first part is too hard for him. then the system will adjust based on his quiz data and interaction with system duiring study session and also the intial four dimential perefernce.

\subsubsection{\texttt{AdaptMate}: Intelligent Plan Adaptation}
\label{sec:adaptmate}

\begin{figure*}[!tb]
  \includegraphics[width=\textwidth]{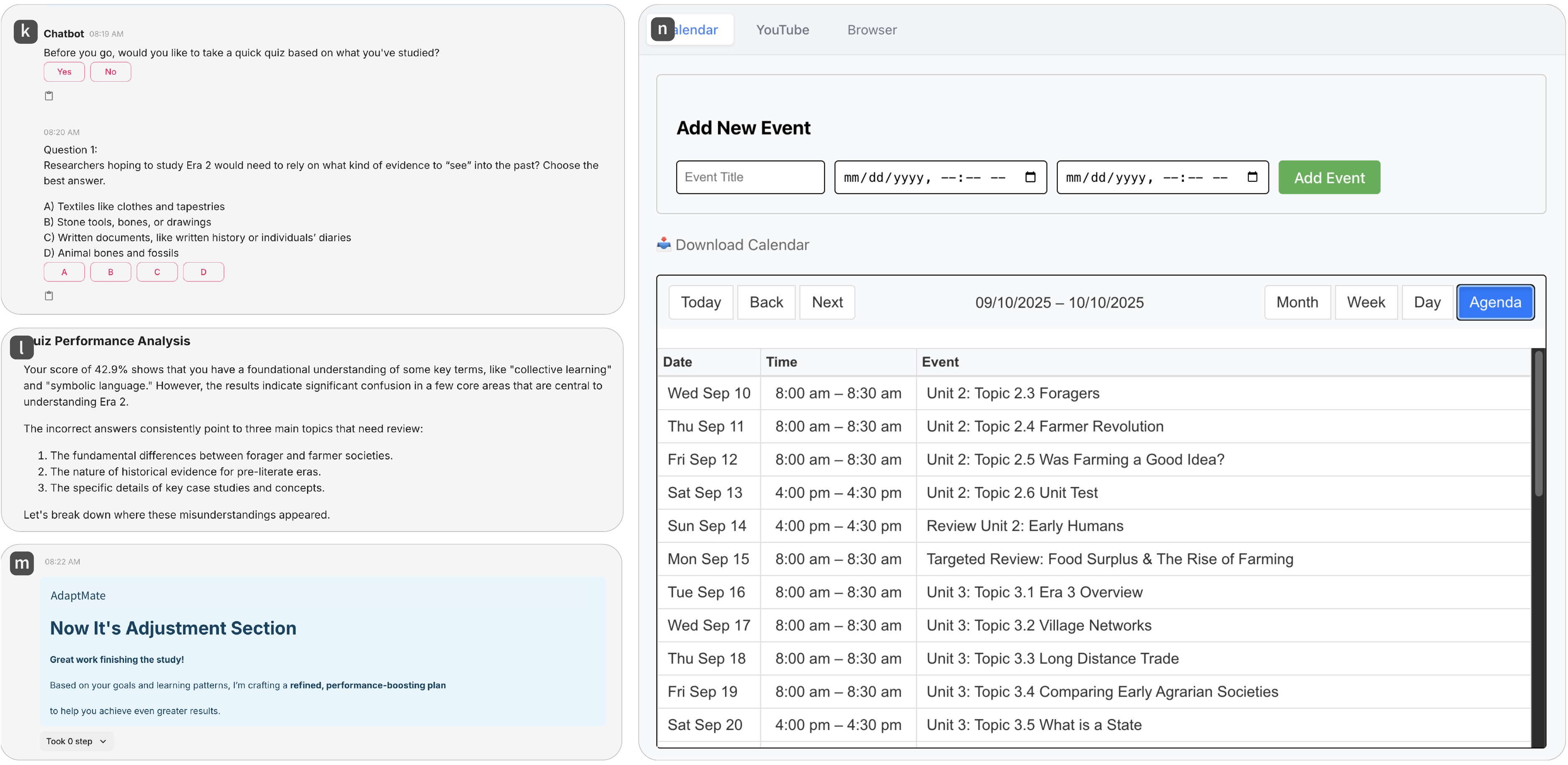}
   \vspace{-12pt}
  \caption{\textit{\texttt{AdaptMate} Interface ---} The front-end interface for adaptive plan refinement based on learning performance. After completing a study session and quiz, users receive a detailed performance report with learning insights (step \protect\darkgreysquare{l}). Users can then request plan adjustments (step \protect\darkgreysquare{m}), and \texttt{AdaptMate} generates an updated study plan that incorporates quiz results, interaction patterns, and prior preferences, presented in both text and calendar formats (step \protect\darkgreysquare{n}). We outline the user's interaction with the front-end as a guide to explain the pipeline of \texttt{AdaptMate} in Section \ref{sec:adaptmate}.}
  \label{fig:adapt_system}
  \Description[AdaptMate Interface]{A split-screen screenshot. The left panel shows a chat interface with three labeled regions. (k) A quiz question asking about evidence historians use to study Era 2, with four answer options (A–D) and clickable answer buttons. (l) A quiz performance analysis showing a score of 42.9\%, identifying three areas of confusion: differences between forager and farmer societies, nature of historical evidence in pre-literate eras, and specific details of key case studies. (m) An AdaptMate section titled "Now It's Adjustment Section" offering to generate a refined, performance-boosting plan. The right panel (n) shows an agenda-view calendar for September 10–October 10, 2025, listing adapted study sessions including targeted review sessions such as "Targeted Review: Food Surplus \& The Rise of Farming."}
\end{figure*}

The front-end interface of \texttt{AdaptMate} is shown in Figure~\ref{fig:adapt_system}.
After finishing the quiz, the system automatically generates a comprehensive quiz report with performance analysis and future learning suggestions for Alex (step \darkgreysquare{l}). 
Following their study session and quiz completion, Alex can request plan adjustments by clicking \textit{Generate Adjusted Plan} if they find certain content too challenging or wish to modify their learning approach (step \darkgreysquare{m}). 
\texttt{AdaptMate} analyzes their quiz performance, interaction patterns during the study session, and original four-dimensional preferences to propose modifications. The updated plan is presented in both text and calendar format, maintaining consistency with the original \texttt{PlanMate} interface (step \darkgreysquare{n}).
% The system presents recommended changes such as additional time allocation for difficult topics, alternative content sequences, or adjusted pacing, which Alex can review and approve before implementation.

\paragraph{Technical Details. } 
% This agent provides targeted feedback on performance patterns and generates specific recommendations for future learning sessions based on identified knowledge gaps and strengths.
This agent analyzes performance patterns to provide targeted feedback and recommendations based on identified knowledge gaps and strengths.

\texttt{AdaptMate} analyzes multiple data sources to determine optimal plan adjustments: quiz performance results from \texttt{StudyMate}, user interaction patterns during study sessions (question frequency, content areas where questions were asked, session completion rates), content coverage tracking, and the user's original four-dimensional preferences from \texttt{PlanMate}. 
The system uses structured prompts to evaluate this combined data and generate specific recommendations such as additional time allocation for challenging topics, alternative content sequences, or adjusted pacing preferences. 
These recommendations are then processed through the same planning generation pipeline as \texttt{PlanMate} to support schedule feasibility and maintain learning progression coherence.

\subsection{Implementation Details}
The LLM agents in the backend of \tool{} are powered by Gemini-2.0-Pro \cite{gemini2025}, and the front-end is implemented using React \cite{react}. Agent prompts and source code are available in the supplementary materials. \footnote{The supplementary materials can be found at \url{https://osf.io/z9guh/?view_only=46f35fd0b2bd4794a1e92ce6e86160b2}}

\section{Preliminary User Studies}% for \texttt{PlanMate} \& \texttt{StudyMate}}
\subsection{Method}
% {\color{blue}{This preliminary study aimed to assess the feasibility and usability of \tool{}'s core functionalities, \texttt{PlanMate} and \texttt{StudyMate}, and to identify design refinements for the final user study.
% We conducted an exploratory within-subjects design with two scenarios: \textit{S1 (Personalized Study Plans)} and \textit{S2 (Real-time Contextual Assistance)}. In each scenario, participants experienced both a \tool{} condition (C1) and a baseline condition (C2: Khan Academy with Gemini-2.5-pro \cite{khanacademy, gemini2025}) in counterbalanced order. 
% S2 participants also completed ten-question multiple-choice quizzes after each condition. All sessions ended with semi-structured interviews. 
% The study lasted approximately one hour in total and participants were compensated \$13/hour upon completion. Quizzes and interview protocols are provided in the supplementary materials. 
% \footnote{Supplementary materials: \url{https://osf.io/z9guh/?view_only=46f35fd0b2bd4794a1e92ce6e86160b2}} 
% 24 participants (ages 19--78, $M=36.875$, $SD=18.087$) were recruited via university mailing lists; 87.5\% had prior online learning experience and 70.8\% were native English speakers. Participants are referenced as P1--P24.
% We conducted paired t-tests on quiz scores ($\alpha=0.05$) and applied Thematic Analysis to interview transcripts~\cite{clarke2014thematic, McDonald19}.}}

% Old Method
% -------------------------------------------------------------------------------
\subsubsection{Study Design}
% This study aimes to undersrtadn the imapct of \tool{} on leaner's learning outcomes and user experience,  specifically evalute how well personalzie planning, realtime contextual assitance can impact leanring outcome and user experience and  identify refinements for the final user study design. We conduct different ablation condtions served as the wthin-subjects varible. 
% particianet are seperated into two groups for two scenarios. secnario 1 is planning, sceanrio 2 is sudying. below we describe contdtions in two scenarios design.
This preliminary study aimed to assess the feasibility and usability of \tool{}'s core functionalities and to identify design refinements for our final user study.
This study focused on evaluating two of \tool{}'s components: \texttt{PlanMate}'s personalized study plans and \texttt{StudyMate}'s real-time contextual assistance, and their impact on learning outcomes and experience.

We conducted an exploratory within-subjects design with two scenarios, each targeting one of the \tool{}'s core functionalities. Participants were randomly separated into two groups, and each group experienced one scenario that contained two conditions. %Below we describe the two scenario designs and the conditions.
For clarity, we refer to the scenarios as \textit{S1 (Personalized Study Plans)} and \textit{S2 (Real-time Contextual Assistance)}, and to the conditions as \textit{C1 (\tool{})} and \textit{C2 (Baseline)}. 
% {\color{blue}
% {The baseline was chosen to reflect how students typically supplement online learning in practice, using a general-purpose LLM alongside an online course platform.}}
The baseline was chosen to reflect how students typically supplement online learning in practice, using a general-purpose LLM alongside an online course platform.

\paragraph{Scenario 1: Personalized Study Plans}
In this scenario, participants were asked to design and create learning plans for \textit{World History Project} courses from Khan Academy \cite{khanacademy}. 
In Condition 1, participants engaged with \tool{}'s \texttt{PlanMate} feature. They designed and created personalized learning plans for equivalent course content using \tool{}.
In Condition 2, participants used Khan Academy courses with Gemini-2.5-pro \cite{gemini2025} assistance to design and create a learning plan for \textit{World History Project} courses.
% In Condition 2, participants experienced standard online learning with LLM support using Khan Academy courses and Gemini-2.5-pro \cite{gemini2025} for planning assistance. They were asked to design and create a learning plan for \textit{World History Project} courses.

\paragraph{Scenario 2: Real-time Contextual Assistance}
In this scenario, participants were asked to study course content from Khan Academy's \textit{World History Project} for 20 minutes. Course materials included \textit{Era 3 Overview} and \textit{Era 2 Overview} from Khan Academy's World History Project. 
In Condition 1, participants engaged with our system's \textit{real-time contextual assistance} feature. 
In Condition 2, participants experienced standard online learning with Gemini-2.5-pro as general study assistance. \\

% Within each scenario, participants experienced both conditions in counterbalanced order, with course materials also counterbalanced to mitigate order effects. 
Within each scenario, participants experienced both conditions in counterbalanced order, with course materials also counterbalanced to mitigate order effects and reduce the influence of participants' prior knowledge of the materials and individual learning strategies on learning outcomes.
For S2, participants completed knowledge assessment quizzes after each condition. At the end of each scenario, we conducted semi-structured interviews to gather qualitative feedback on user experience and identify opportunities for system improvement.
The study lasted approximately one hour in total and participants were compensated \$13/hour upon completion. Quizzes and interview protocols are provided in the supplementary materials. \footnote{The supplementary materials can be found at \url{https://osf.io/z9guh/?view_only=46f35fd0b2bd4794a1e92ce6e86160b2}}

% for consstency, we denoted scaarios and condtions with S1(Personalized Study Plans scenaio), S2(Real-time Contextual Assistance scenaio), C2(codntion 2), C1(Condition 1 ).
% During the study, participants wee ranomly ssigned to both condtions and both courses/untis in a radomzied order. After each contion, prtiicipants compeleted the quantiviave scales. at the end of thier interation with each scenario, semi-seturcture interviews were conducted. the entire study lasrte 1 hour. questionarries userd duing the study can be foudn in the supplmeatry materials.

% We design 2 scearios to compare using our system, with state-of-the-art online learning platform Khan academy and an LLM gemini for learning support \cite{khanacademy, gemini2025}

\subsubsection{Measures}
% To evaluate the participants' learning outcomes, we selected and employed and created two knowledge assessment quizzes sepeartely from \textit{3.1 Era 3 Overview} and \textit{2.1 Era 2 Overview} from Khan Academy.
% Both quizess were 4 answers mcq?
To evaluate participants' learning outcomes, we selected and created two knowledge assessment quizzes separately from \textit{Era 3 Overview} and \textit{Era 2 Overview} from Khan Academy. The quiz questions were adapted from Khan Academy's assessment materials to ensure content validity and appropriate difficulty levels for the target learning objectives. Both quizzes consisted of ten four-answer multiple-choice questions designed to assess participants' comprehension and retention of the course material. 

\subsubsection{Participants}
24 participants were recruited for our user study. Participants were required to be in the United States, fluent in English, and at least 18 years old. All participants were recruited through university mailing lists. 
While our sample size is not large, the within-subjects study design achieves an acceptable level of statistical power for significant results \cite{bellemare2014statistical}.
Participants' ages ranged from 19--78 ($M = 36.875$, $SD = 18.087$). The majority of participants (87.5\%) reported prior learning experience with online learning platforms, while 12.5\% had no such experience. Regarding language background, 70.8\% were native English speakers and 29.2\% were non-native English speakers. Throughout this paper, participants are referenced as P1--P24, where P\textit{i} denotes the \textit{i}th participant.

% to do tommorrow
% Participants age ranged from 19--48 ($M = 25$, $SD = 7.9$). 50\% of the participants identified as female and 50\% as male. 50\% of our participants were White, 41.6\% were Asian, and 8.4\% were American Indian or Alaska Native. After the study, participants were compensated \$15.00 per hour. We refer to participants as P1--P12, using the notation P\textit{i} to indicate participants, where \textit{i} indicates participant ID number.
% \revision{In the recruitment survey, we also collected participants' experiences with LLMs, asking them to select a category that best described their familiarity: ``not familiar or none,'' ``occasional use,'' or ``regular use.'' Five participants (P7--P11) selected ``not familiar or none,'' four (P1, P4, P6, P12) selected ``occasional use,'' and three (P2, P3, P5) selected ``regular use.'' Those who reported occasional or regular use mentioned using LLMs for tasks such as brainstorming, search engines, writing assistance, and planning tools (\eg scheduling assistance, task management, project coordination, and itinerary planning.)}

\subsubsection{Analysis}
% For qualitative data, we conducted a Thematic Analysis (TA) on the interviews. The coding of the responses was conducted by deriving representative themes from transcriptions~\cite{clarke2014thematic, McDonald19}. During open coding, the first author coded for significant concepts in the data. Concepts were then categorized into clusters, further being grouped into themes. These themes were iteratively discussed between the whole research team, recategorizing the groups and revising the themes upon disagreement until a consensus was reached. 

For the quantitative data, we conducted paired t-tests to compare the means between the C2(baseline) and the C1(\tool{}) for quiz scores within S2 (real-time contextual assistance) on both Quiz 1 (\textit{Era 3 Overview}) and Quiz 2 (\textit{Era 2 Overview}). The tests were performed with an alpha level of 0.05.

For qualitative data, we conducted a Thematic Analysis (TA) on the interview transcripts. We followed an inductive coding approach to derive representative themes from the data~\cite{clarke2014thematic, McDonald19}. During initial coding, the first author coded and identified significant concepts within the transcripts. These concepts were subsequently categorized into clusters and grouped into themes. The entire research team engaged in discussions to refine these themes, resolving disagreements through collaborative review and revising the thematic framework until consensus was achieved.

% For qualitative data, we conducted a Thematic Analysis (TA) on the interviews. The coding of the responses was conducted by deriving representative themes from transcriptions~\cite{clarke2014thematic, McDonald19}. During open coding, the first author coded for significant concepts in the data. Concepts were then categorized into clusters, further being grouped into themes. These themes were iteratively discussed between the whole research team, recategorizing the groups and revising the themes upon disagreement until a consensus was reached.

% Our analysis aimed to assess the impact of each feature of our system on online learning and identify refinements for the system.
% The results of our quantitative data are shown in Figure~\ref{fig:quizold}.
% We found that participants in the \tool{} condition ($M=9.167$, $SD=0.408$) scored higher on Quiz 1 than the baseline condition ($M=8.333$, $SD=1.367$), while Quiz 2 scores were slightly lower in the \tool{} condition ($M=5.667$, $SD=1.211$) compared to baseline ($M=6.000$, $SD=1.414$).

% Below, we present our findings in seven key themes that emrged in our analysis. The first six themes talk about the user interaction experiences for features \texttt{PlanMate} and \texttt{StudyMate}. For the seventh theme, we derived from qualitative themes the improvement of the system.
%Both themes are derived from quantitative and qualitative findings.

\subsection{Results} 
Our analysis aimed to assess the effects of individual system features on online learning outcomes and user experiences and identify refinements for the system.
% The results of the quantitative data are shown in Figure~\ref{fig:quizold}.
Quantitatively, participants in the \tool{} condition ($M=9.167$, $SD=0.408$) scored higher on Quiz~1 than those in the baseline condition ($M=8.333$, $SD=1.367$), whereas Quiz~2 scores were slightly lower in the \tool{} condition ($M=5.667$, $SD=1.211$) compared to baseline ($M=6.000$, $SD=1.414$); neither difference reached statistical significance (Quiz~1: $F(1,10)=2.049$, $p=.183$; Quiz~2: $F(1,10)=0.192$, $p=.670$).

% Below, we present seven key themes that emerged from our quantitative and qualitative analysis.
% The first six themes describe users' interaction experiences with the \texttt{PlanMate} and \texttt{StudyMate} features.
% The seventh theme synthesizes participants' feedback to highlight directions for system improvement.

Below, we present seven key themes that emerged from our quantitative and qualitative analysis.
The first six themes describe users' interaction experiences with the \texttt{PlanMate} and \texttt{StudyMate} features.
The seventh theme synthesizes participants' feedback to highlight directions for system improvement.

\subsubsection{\texttt{PlanMate} Supported Learning Instrumentally and Psychologically}
Participants described \texttt{PlanMate} as supporting learning at both instrumental and psychological levels. 
Instrumentally, participants emphasized that \texttt{PlanMate} helped them manage time and structure tasks in ways that made learning more feasible within their everyday lives. Two participants (P8 and P12) noted that the system's personalized planning guidance was particularly helpful for learners juggling multiple responsibilities and time constraints. 
Psychologically, participants reported that \texttt{PlanMate} increased their motivation and sense of autonomy, making learning feel more doable and less overwhelming overall. Five participants (P2, P3, P5, P6, P10) noted that personalized learning plans encouraged them to feel more motivated and organized. Ten participants (P1--P3, P5--P11) further emphasized that \texttt{PlanMate} supported autonomy by enabling learners to create and adjust their own learning plans. Four participants (P5, P6, P9, P10) described how personalized learning plans helped them better align studying with their personal needs and learning goals. 

\subsubsection{\texttt{PlanMate} Enabled More Accurate and Actionable Learning Plans Beyond State-of-the-Art LLMs}
Eight participants (P3, P5--P10, P11) described the personalized learning plans generated by \texttt{PlanMate} as accurate, realistic, and manageable. 
% Participants emphasized that the plans accounted for multiple constraints, such as available time, deadlines, workload, and pacing, resulting in schedules that felt accurate and realistic. 
Five participants (P1, P6--P8, P10) further noted that the baseline approach lacked lookahead and failed to provide actionable learning plans. 

\subsubsection{The Visual Calendar-Based Interface of \texttt{PlanMate} Increased Engagement and Commitment}
Five participants (P2, P3, P5, P7, P9) described \texttt{PlanMate}'s calendar-based visual interface as more engaging and motivating than text-only outputs from existing LLM tools. Participants explained that presenting learning plans in a manipulable calendar view made the plans easier to understand, revisit, and commit to over time. Participants also suggested that interacting with the calendar, such as being able to mark completed tasks, would further help them track progress and maintain motivation.

\subsubsection{\texttt{StudyMate} Provided Contextual Support and a Learning-Suited Interface Beyond State-of-the-Art LLMs}
Participants primarily highlighted two key strengths of \texttt{StudyMate} that they felt were missing from existing state-of-the-art LLM tools such as Gemini. 
First, seven participants (P14, P15, P17, P19, P21, P23, P24) valued \texttt{StudyMate}'s contextual support, including real-time assistance and domain-specific knowledge directly grounded in the study material. 
Second, six participants (P13, P16, P17, P18, P20, P23) emphasized the importance of a more usable, learning-suited interface.

\subsubsection{\texttt{StudyMate} Enabled Teacher-Like Support Through In-Context Features}
Participants primarily described four in-context, real-time support features, specifically video summaries, timestamps, review quizzes, and external resources, as particularly helpful, allowing \texttt{StudyMate} to function as teacher-like support by providing accessible assistance that is often unavailable in isolated online learning settings.
Eight participants (P13--P18, P21, P23) highlighted the video summarization feature as especially useful for reviewing previously learned content and preparing for exams. 
Six participants (P17, P19--P22, P24) emphasized the usefulness of video timestamps for understanding content in context and preparing for exams. 
Two participants (P1 and P10) additionally mentioned that review quizzes and links to external resources supported deeper learning. 
Additionally, five participants (P15--P17, P19, P24) reported that \texttt{StudyMate} made them more willing to ask questions and helped them remember content better.

\subsubsection{\texttt{StudyMate} Deepened Learning and Built Trust Through Contextual Grounding}
Participants described contextual grounding as supporting learning at both a functional and psychological level.
Functionally, grounding AI responses in the study material made confusing concepts easier to understand and enabled deeper learning. 
Nine participants (P13, P14, P17--P23) reported that contextual grounding helped them better understand confusing parts of the material and achieve deeper comprehension. 
Psychologically, contextual grounding increased trust in the system's answers, which is critical for building rapport with the AI and sustaining long-term use for learning. One participant (P15) emphasized that contextual grounding increased their trust in the AI's responses.

\subsubsection{Improving \texttt{PlanMate} and \texttt{StudyMate}: Motivating the Need for \texttt{AdaptMate}}

While quantitative results showed directional differences between conditions, these effects did not reach statistical significance. Our qualitative findings help contextualize these results and reveal two primary directions for improving \tool{} that motivate the need for \texttt{AdaptMate}, as well as one study design implication. Below, we first present system-level improvement directions, followed by the study design improvement.

% \textbf{System-Level Improvements for Adaptive Mixed-Initiative Learning.}
Participants explicitly identified two primary areas of improvement for adaptive mixed-initiative learning.
First, ten participants (P13--P18, P20, P21, P23, P24) expressed a preference for receiving concise, domain-specific answers grounded in the course material, supplemented by optional external resources. Participants reported frustration when responses were overly long or insufficiently specific to the learning context. They valued being able to selectively explore additional information when needed, rather than being presented with extensive explanations upfront. At the same time, access to broader external knowledge was viewed as beneficial for supporting deeper or more comprehensive learning.
As P17 explained:
\textit{``General answers are frustrating to me, but for your system, I kept pressing the `more details' button because I like to pick and choose what I want to learn. I would prefer it to be more specific to the content I'm consuming, but also have access to external sources when needed.''}
Second, five participants (P5, P7--P9, P12) emphasized the importance of being able to adapt learning plans over time as their schedules or circumstances changed. Participants noted that fixed plans quickly become outdated, and that repeatedly re-prompting the system to generate entirely new plans was inefficient. Instead, they preferred incremental adjustments that allow plans to evolve while preserving prior structure.
As P9 explained:
\textit{``I think that the ability to align it with my needs is good, and that being able to go back and adjust it as needed is really important.''}

% \textbf{Study Design and Assessment Improvement.}  
Findings also suggest an assessment-related study design improvement. The lack of a significant difference in Quiz~1 scores may reflect a ceiling effect, as the baseline median score was $8.33$ out of 10, leaving limited room for measurable improvement. To address this limitation, future iterations of the study will include additional, more challenging questions to better differentiate learning outcomes. 
For Quiz~2, the slightly lower performance in the \tool{} condition may be attributed to questions that extended beyond the scope of the instructional video, requiring external knowledge or reasoning not explicitly supported by the system. As P22 noted:
\textit{``There were questions that the video didn't seem to cover in any detail, as far as I could tell.''}

These findings motivated both system-level and study design refinements. 
Specifically, participant feedback highlighted the need for adaptive plan refinement during learning, improved support for selectively expanding beyond course materials, and a more sensitive assessment of learning outcomes.

\begin{figure}[!tb]
  \includegraphics[width=\columnwidth]{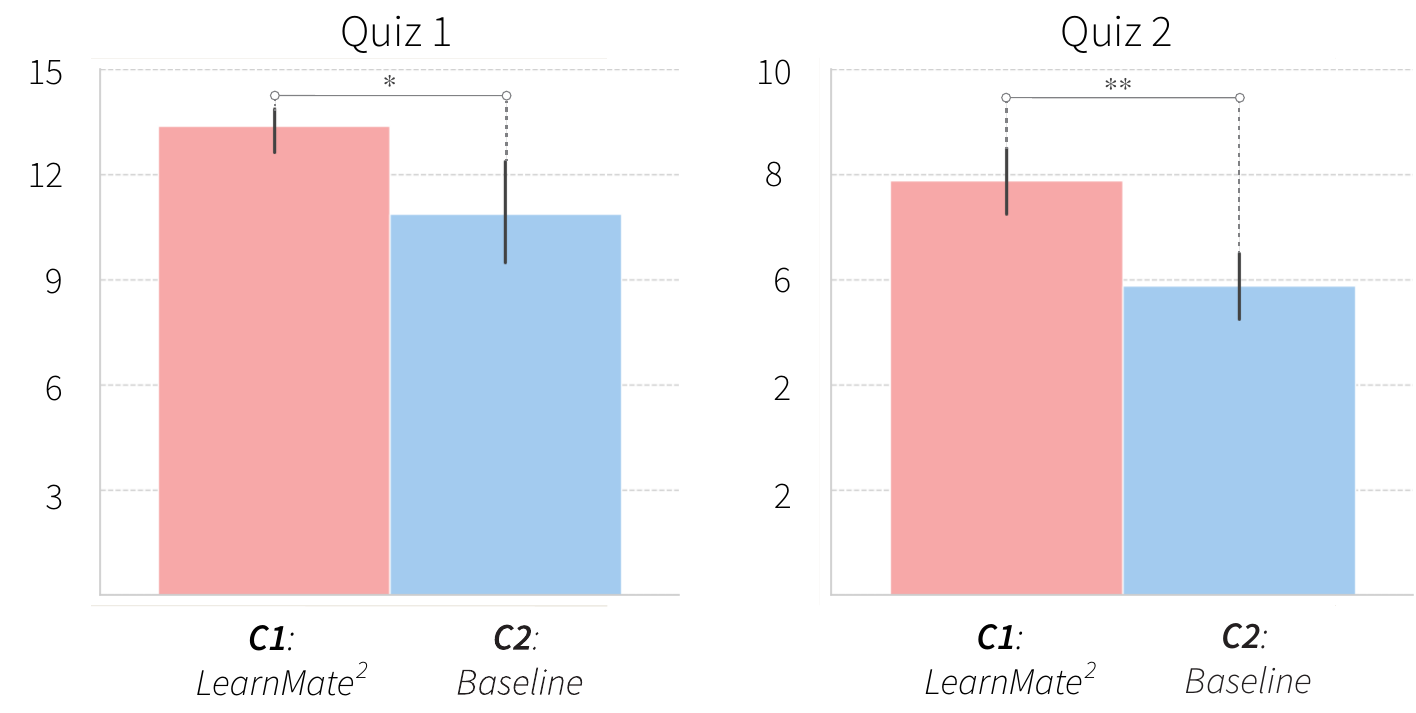}
   \vspace{-12pt}
  \caption{\textit{Quantitative Data from User Study---}
  Bar graphs show participants' quiz results for Quiz 1 and Quiz 2. Horizontal lines indicate significant pairwise comparisons with paired t-tests ($p < .05^{\ast}$, $p < .01^{\ast\ast}$, $p < .001^{\ast\ast\ast}$). Vertical lines in each bar graph indicate standard error.}
  \label{fig:quiznew}
  \Description[Quiz Results]{Two side-by-side bar charts comparing quiz scores between C1 (LearnMate², shown in pink) and C2 (Baseline, shown in blue). Quiz 1 (out of 15): LearnMate² scored approximately 13, Baseline approximately 11, with a statistically significant difference (p < .05, marked with *). Quiz 2 (out of 10): LearnMate² scored approximately 7.9, Baseline approximately 6, with a statistically significant difference (p < .01, marked with **). Error bars indicate standard error.}
\end{figure}

\section{User Study for \tool{}}
\subsection{Method}
\subsubsection{Study Design}
Building on the findings from the preliminary study and the qualitative insights discussed above, we implemented three major system enhancements: (1) added five additional challenging questions to Quiz~1 to address the ceiling effect, (2) integrated external resources with limited scope in \texttt{StudyMate} to provide broader context while maintaining course focus, and (3) the introduction of \texttt{AdaptMate} to support formative adaptation of personalized plans.
This study aims to assess the impact of the integrated workflow on learning outcomes and user experience.

We conducted a comparative study using a within-subjects design. In Condition 1, participants engaged with the complete \tool{} system, which integrated personalized study plan guidance and generation, real-time contextual assistance during learning, and adaptive learning activities that adjust based on ongoing progress assessment.
In Condition 2, participants experienced standard online learning with assistance from Gemini-2.5-pro and self-directed planning, learning and adaptation strategies.
In both conditions, participants engaged in a complete learning cycle with three tasks. In the first task, they created a learning plan for the entire \textit{``World History Project''} course from Khan Academy. In the second task, they planned their specific study session and studied according to their plan for 30 minutes (studying). In the final task, they adapted their plan based on their learning experience and progress (adaptation). 
% {\color{blue}{More specifically, in the \tool{} condition, participants completed T3 using \texttt{AdaptMate}, which proposed plan modifications based on their quiz performance, in-session interactions, and original preferences. Participants then reviewed and accepted, modified, or rejected the changes. In the baseline condition, participants manually revised their plan using Gemini-2.5-pro based on their quiz results.}}
More specifically, in the \tool{} condition, participants completed T3 using \texttt{AdaptMate}, which proposed plan modifications based on their quiz performance, in-session interactions, and original preferences. Participants then reviewed and accepted, modified, or rejected the changes. In the baseline condition, participants manually revised their plan using Gemini-2.5-pro based on their quiz results.
Course materials included \textit{``Era 2 Overview''} and \textit{``Era 3 Overview''} from Khan Academy's World History Project.
For consistency, we denote these conditions and tasks as \textit{C1 (\tool{})}, \textit{C2 (Baseline)}, \textit{T1 (Planning)}, \textit{T2 (Studying)}, and \textit{T3 (Adaptation)} in the remainder of the paper. In the \tool{} condition, T1, T2, and T3 were supported by \texttt{PlanMate}, \texttt{StudyMate}, and \texttt{AdaptMate}, respectively, while the baseline condition used standard, non-integrated learning tools to reflect how students commonly supplement online learning in practice.
% {\color{blue}{to reflect how students commonly supplement online learning in practice.}}

During the study, participants were randomly assigned to both conditions within their scenario, with course materials and condition order counterbalanced. 
% Within each scenario, participants experienced both conditions in counterbalanced order, with course materials also counterbalanced to mitigate order effects and reduce the influence of participants' prior knowledge of the materials and individual learning strategies on learning outcomes.
After each task, participants completed quantitative scales specific to the task they had just completed. Enhanced knowledge assessment quizzes were administered specifically after T2 (studying).
At the end of their interaction with both conditions, semi-structured interviews were conducted to gather qualitative insights about user experience and system effectiveness.
The entire study lasted approximately 1.5 hours and participants were compensated \$13/hour upon completion. Questionnaires, quizzes, and interview protocols used during the study can be found in the supplementary materials.
\footnote{The supplementary materials can be found at \url{https://osf.io/z9guh/?view_only=46f35fd0b2bd4794a1e92ce6e86160b2}} 

\begin{figure*}[!tb]
  \includegraphics[width=\textwidth]{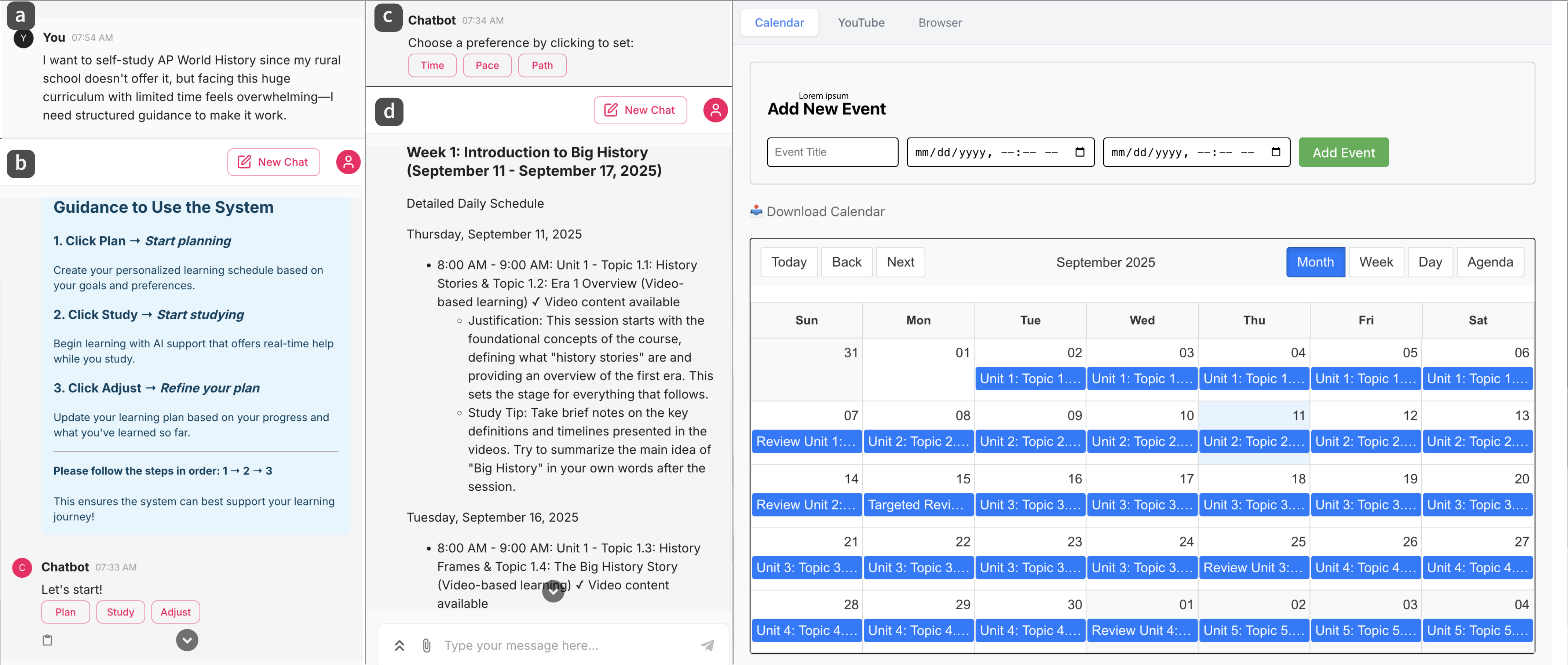}
   \vspace{-12pt}
  \caption{\textit{Quantitative Data from User Study ---} Bar graphs on participants' perceived performance of usefulness, ease of use, ease of learning, satisfaction and usability scores across different conditions for the planning task (T1). Horizontal lines indicate significant pairwise comparisons with repeated measures ANOVA ($p < .05^{\ast}$, $p < .01^{\ast\ast}$, $p < .001^{\ast\ast\ast}$). Vertical lines in each bar graph indicate standard error.}
  \label{fig:plan}
  \Description[Planning Task Results]{Five side-by-side bar charts comparing C1 (PlanMate, pink) and C2 (Baseline, blue) on the planning task (T1) across five measures, all on a 1–7 scale except SUS which is 0–100. Performance: PlanMate approximately 5.5, Baseline approximately 4.5, no significant difference. Ease of Use: both approximately 5.5, no significant difference. Ease of Learning: both approximately 6, no significant difference. Satisfaction: PlanMate approximately 5.5, Baseline approximately 4, no significant difference. SUS: PlanMate approximately 75, Baseline approximately 65, no significant difference. Error bars indicate standard error.}
\end{figure*}

\subsubsection{Measures}
% For qualitative data, we conducted a Thematic Analysis (TA) on the interview transcripts. We followed an inductive coding approach to derive representative themes from the data~\cite{clarke2014thematic, McDonald19}. During initial coding, the first author coded and identified significant concepts within the transcripts. These concepts were subsequently categorized into clusters and grouped into themes. The entire research team engaged in discussions to refine these themes, resolving disagreements through collaborative review and revising the thematic framework until consensus was achieved.

To evaluate participants' experiences with the system, we employed the Usefulness, Satisfaction, and Ease of use (USE) scale~\cite{gao2018psychometric} to measure four key dimensions: usefulness (Cronbach's $\alpha$: T1 = $0.96$, T2 = $0.97$, T3 = $0.96$), ease of use (Cronbach's $\alpha$: T1 = $0.91$, T2 = $0.94$, T3 = $0.92$), ease of learning (Cronbach's $\alpha$: T1 = $0.90$, T2 = $0.94$, T3 = $0.94$), and satisfaction (Cronbach's $\alpha$: T1 = $0.96$, T2 = $0.97$, T3 = $0.97$).
We also employed the System Usability Scale (SUS)~\cite{brooke1996sus} to assess overall system usability (Cronbach's $\alpha$: T1 = $0.78$, T2 = $0.87$, T3 = $0.80$). 
The USE scales were administered on a seven-point Likert scale, while SUS used a five-point scale.

To evaluate participants' learning outcomes, we selected and created two knowledge assessment quizzes based on \textit{``Era 3 Overview''} and \textit{``Era 2 Overview''} from Khan Academy. 
Compared to the preliminary study, Quiz 1 was enhanced with five additional challenging four-answer multiple-choice questions that extend beyond video content, requiring external knowledge and reasoning. These questions were carefully designed to address the ceiling effect observed in the original assessment. The first ten questions remained unchanged from the preliminary study.
Quiz 2 remained unchanged. 
Quiz 1 consisted of 15 multiple-choice questions with four options each, while Quiz 2 contained 10 questions.

\subsubsection{Participants}
16 participants were recruited for our user study. Participants were required to be in the United States, fluent in English, and at least 18 years old. All participants were recruited through university mailing lists. None of the participants in this user study participated in the preliminary study.
While our sample size is not large, the within-subjects study design achieves an acceptable level of statistical power for significant results \cite{bellemare2014statistical}.
Participants' ages ranged from 19--44 ($M = 26.25$, $SD = 8.299$). The majority of participants (93.7\%) reported prior learning experience with online learning platforms, while 6.3\% had no such experience. Regarding language background, 75\% were native English speakers and 25\% were non-native English speakers. Participants are referenced as P1--P16, where P\textit{i} denotes the \textit{i}th participant.

\subsubsection{Analysis}
For the quantitative data, we conducted repeated measures ANOVA to compare the means between the \tool{} condition (C1) and the baseline condition (C2) for USE and SUS scores across all three tasks (T1, T2, T3), as well as for quiz scores on both Quiz 1 (\textit{Era 3 Overview}) and Quiz 2 (\textit{Era 2 Overview}). 
% {\color{blue}{We chose repeated measures ANOVA over paired t-tests to account for the within-subjects design and to simultaneously compare conditions across multiple tasks (T1, T2, and T3).}}
We chose repeated measures ANOVA over paired t-tests to account for the within-subjects design and to simultaneously compare conditions across multiple tasks (T1, T2, and T3).
All tests were performed with an alpha level of 0.05.

For qualitative data, we conducted a Thematic Analysis (TA) on the interview transcripts. We followed an inductive coding approach to derive representative themes from the data~\cite{clarke2014thematic, McDonald19}. During initial coding, the first author coded and identified significant concepts within the transcripts. These concepts were subsequently categorized into clusters and grouped into themes. The entire research team engaged in discussions to refine these themes, resolving disagreements through collaborative review and revising the thematic framework until consensus was achieved.

\begin{figure*}[!b]
  \includegraphics[width=\textwidth]{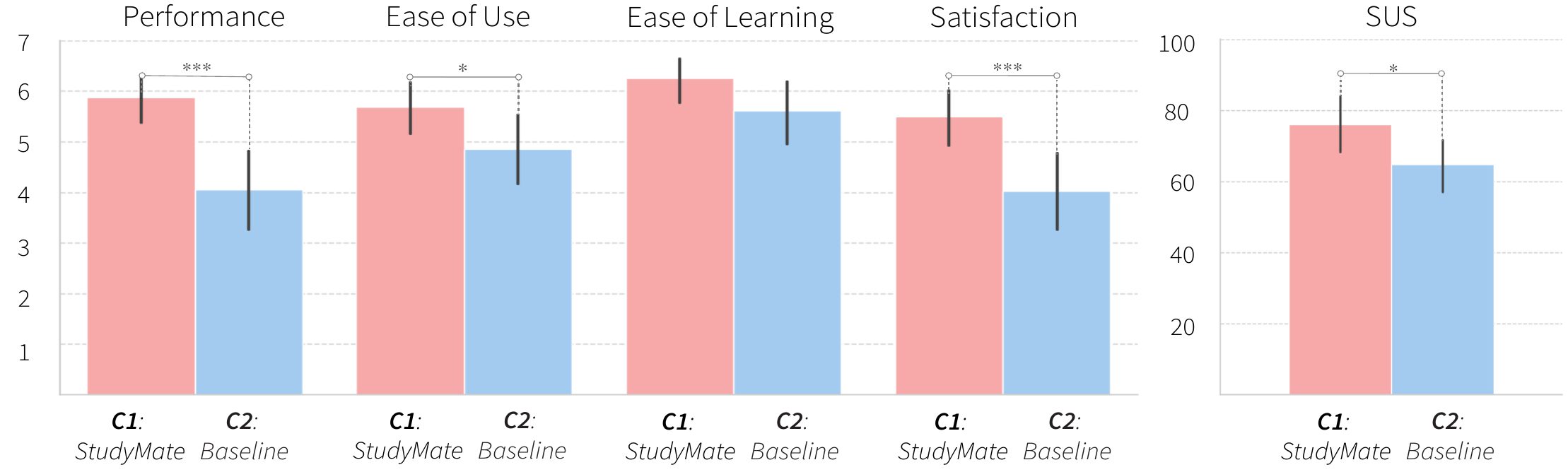}
   \vspace{-12pt}
    \caption{\textit{Quantitative Data from User Study---} Bar graphs on participants' perceived performance of usefulness, ease of use, ease of learning, satisfaction and usability scores across different conditions for the studying task (T2). Horizontal lines indicate significant pairwise comparisons with repeated measures ANOVA ($p < .05^{\ast}$, $p < .01^{\ast\ast}$, $p < .001^{\ast\ast\ast}$). Vertical lines in each bar graph indicate standard error.}
  \label{fig:study}
\Description[Studying Task Results]{Five side-by-side bar charts comparing C1 (StudyMate, pink) and C2 (Baseline, blue) on the studying task (T2). Performance: StudyMate approximately 5.9, Baseline approximately 4.1, significant difference (p < .001, marked ***). Ease of Use: StudyMate approximately 5.7, Baseline approximately 4.9, significant difference (p < .05, marked *). Ease of Learning: StudyMate approximately 6.3, Baseline approximately 5.6, no significant difference. Satisfaction: StudyMate approximately 5.5, Baseline approximately 4.0, significant difference (p < .001, marked ***). SUS: StudyMate approximately 76, Baseline approximately 65, significant difference (p < .05, marked *). Error bars indicate standard error.}
\end{figure*}

\subsection{Results} 
\label{userstudy_findings} 

Our analysis examined the impact of the improved \tool{} on learners' experiences and learning outcomes. We investigated how large language models (LMs) supported online learning in practice, what aspects of the system participants found helpful or challenging, and how specific system features shaped learners' interactions, understanding, and planning behaviors. 

\begin{figure*}[!tb]
  \includegraphics[width=\textwidth]{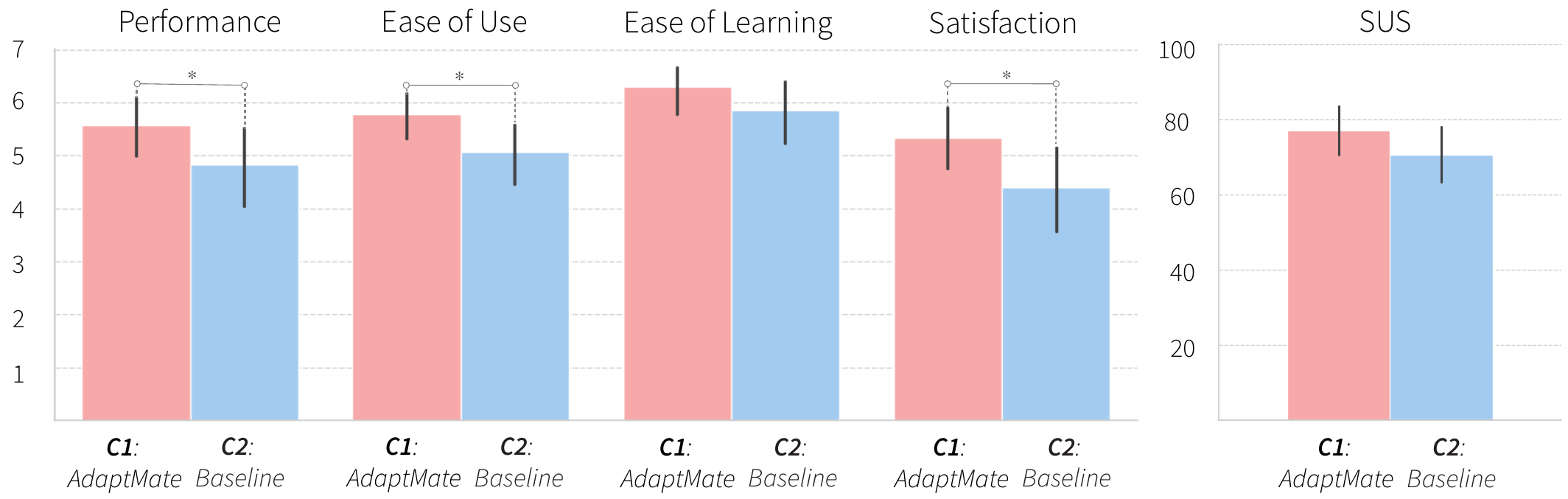}
   \vspace{-12pt}
  \caption{\textit{Quantitative Data from User Study---} Bar graphs on participants' perceived performance of usefulness, ease of use, ease of learning, satisfaction and usability scores across different conditions for the adapting task (T3). Horizontal lines indicate significant pairwise comparisons with repeated measures ANOVA ($p < .05^{\ast}$, $p < .01^{\ast\ast}$, $p < .001^{\ast\ast\ast}$). Vertical lines in each bar graph indicate standard error.}
  \label{fig:adapt}
  \Description[Adapting Task Results]{Five side-by-side bar charts comparing C1 (AdaptMate, pink) and C2 (Baseline, blue) on the adapting task (T3). Performance: AdaptMate approximately 5.6, Baseline approximately 4.8, significant difference (p < .05, marked *). Ease of Use: AdaptMate approximately 5.7, Baseline approximately 5.1, significant difference (p < .05, marked *). Ease of Learning: AdaptMate approximately 6.3, Baseline approximately 5.8, no significant difference. Satisfaction: AdaptMate approximately 5.3, Baseline approximately 4.4, significant difference (p < .05, marked *). SUS: AdaptMate approximately 77, Baseline approximately 71, no significant difference. Error bars indicate standard error.}
\end{figure*}

Our quantitative results are presented in Figure~\ref{fig:quiznew} and Figures~\ref{fig:plan}, \ref{fig:study}, and \ref{fig:adapt}. 
For user experience, we found that \texttt{StudyMate} ($p<.0001$) and \texttt{AdaptMate} ($p=.0452$) were significantly useful, and \texttt{StudyMate} ($p=.0103$) and \texttt{AdaptMate} ($p=.025$) were also significantly easy to use. Participants were significantly satisfied with \texttt{PlanMate} ($p=.0462$), \texttt{StudyMate} ($p=.0007$), \texttt{AdaptMate} ($p=.0143$). 
SUS scores were also significantly higher for \texttt{StudyMate} ($p=.0148$), indicating significantly higher overall usability.
For learning outcomes, we found significant effects in both Quiz 1 ($F(1, 14) = 8.187$, $p=.0126$) and Quiz 2 ($F(1, 14) =16.291$, $p=.0012$). A paired t-test showed that participants in the \tool{} condition (Quiz 1: $M=13.375$, $SD=1.061$; Quiz 2: $M=7.875$, $SD=0.991$) scored significantly higher than those in the baseline condition (Quiz 1: $M=10.875$, $SD=2.232$; Quiz 2: $M=5.875$, $SD=0.991$).

% Below, we report qualitative findings organized into four system-level themes that capture participants' experiences with \texttt{PlanMate} and \texttt{StudyMate}, as well as their preferences for learning support design.
% While these quantitative results demonstrate the effectiveness of the improved \tool{} in terms of usability, satisfaction, and learning outcomes, they do not fully explain how and why participants experienced these benefits in practice. 
% Below, we report qualitative findings organized into four system-level themes that capture participants' overall experiences with \tool{}, as well as their preferences and expectations for personalized, AI-driven educational tools.

Below, we report qualitative findings organized into four system-level themes that capture participants' overall experiences with \tool{}, as well as their preferences and expectations for the design of personalized, AI-driven educational tools.

% To further explain our results, we present our qualitative themes below. The first theme discusses the overall impact of \tool{} on the user experience and learning outcomes. The following three themes address the specific, unique impact of each component. %In the first four themes, we first present quantitative findings, followed by qualitative analysis for explanation and deeper understanding.

\subsubsection{Structured Guidance Makes Learning More Accessible, Manageable, and Actionable}
Participants consistently described structured guidance for navigating and implementing personalized planning and for executing the learning process as making online learning more accessible, manageable, and actionable, particularly for learners who struggle to plan or articulate their needs independently.
Ten participants (P2--P4, P6, P10--P15) reported that \tool{} made learning more accessible by reducing cognitive effort and offloading planning-related thinking. Participants explained that \tool{} helped them clarify what to do next, saved effort in figuring out how to begin, and lowered barriers to engagement. They noted that structured prompting and outputs made the learning process more accurate and easier to start, while interactive questions made learning more engaging. 
% Paritcipants also highlighted that \tool{} was especially helpful for non-native speakers or learners dealing with accented instruction, as it supported clearer understanding.
Participants also highlighted \tool{}'s value for non-native speakers and learners dealing with accented instruction, supporting clearer understanding.
As P14 explained:
\textit{
``The instructions Gemini gave were very vague, so I didn't really know what I was supposed to do or what I was supposed to ask. Your system was much more clear about what it was doing, and it was doing a lot of that thinking for you, which was great. I think that's what students want---doing the thinking for me. It felt much more guided and clear about the goal.''}

% Participant mainly find structured guidance for how to navigate and implement the personalized planning process, and execute the learning process makes online learning more accessible, manageable, and actionable—especially for learners who struggle to plan or articulate their needs on their own.
% Ten participant (P2--P4, P6, P10--P15) mentioned \tool{} makes learning mroe accessible by doning a lot of thinking for you. They expalined LearnMate save a lot of effort; helps non-native speakers or instructors with accent; interactive questions makes learning more fun. also participate noted structured prompting and output makes the whole learning process more accurate and easier to start.

Four participants (P2, P6, P10, P13) described \tool{} as making learning more manageable by breaking plans into concrete, structured steps. Participants explained that decomposing plans reduced back-and-forth adjustments, made the overall learning process feel more organized and less overwhelming, and helped them stay focused. Structured prompts and outputs contributed to more consistent and specific plans, while quiz-based adaptations further refined plans and made them easier to follow over time.
As P6 noted:
\textit{
``It felt much more structured and actually built for what I was trying to do. When I asked Gemini to make me a plan, it felt like it just gave vague study tips. I didn't realize that would be a problem with a normal generative AI, and it really made me appreciate having a structured study plan.''}

% Four participant (P2, P6, P10, P13) mentioned \tool{} makes learning mroe manageable. They expalined \tool{} breaking down the plan makes learning more interesting to learners and manageable; structured prompting and output is needed to avoid learners go back and forth, and makes the plan more conssistent and specific; AM makes adapation on the quiz makes the plan more detailed; easier to follow

Ten participants (P1, P3, P5, P6, P8, P10--P14) reported that \tool{} made learning more actionable by helping them locate specific learning materials and translate high-level goals into concrete actions. Participants explained that topic-specific knowledge allowed the system to surface relevant content directly, supporting more targeted learning than generic support services like McBurney. They also noted that \tool{} supported a mixed mode of learning by helping identify weaknesses and underlying gaps in understanding—activities that learners often do not initiate or are unable to perform on their own.
As P10 explained:
\textit{
``I've used learning skills support through the McBurney Center, and this felt similar, but with much more awareness of the actual courses I'm taking. The person I work with doesn't always know the details of my classes, so having something that can see what I'm learning and walk me through it would be really helpful.''}

% Ten participant (P1, P3, P5, P6, P8, P10--P14) mentioned \tool{} makes learning mroe actionable. They explained LM with topic specific knowledge helps locate the specific learning material; more knowledge than McBurney Center. They also described LM provide mixed way of learning and help recongnize weaknesses, helping learners find the underly part of the learning material, which is something learners won't do or cannot do by themselves.

\subsubsection{\tool{} Extend Learning Support Beyond Online Learners to the Broader Learning Ecosystem}
Participants described how \tool{} supported learning beyond individual online learners, extending its value across learning contexts, roles, and stages. Specifically, they highlighted three ways in which \tool{} expanded the learning ecosystem: supporting learning beyond online settings, assisting instructors, and enabling learners to better memorize content and prepare for exams.

Three participants (P6, P10, P12) noted that \tool{} could support learning beyond purely online contexts, including planning for in-person classes. They explained that context-aware planning and guidance were equally valuable for on-site courses, where learners still need help organizing study time and understanding workload. Participants described \tool{} as making learning support more accessible—similar to having instructor-like guidance outside of class—while noting that existing LLM tools such as Gemini were less helpful for supporting course completion.
As P10 explained:
\textit{
``While going through this, I kept thinking that I wish I had a version of this for my actual classes. I personally have a hard time scheduling study sessions and planning how much time things will take, so having something connected like this would really help.''}

% Participants mentioned that beyond supporting online learners in three dimentions.
% First, three participants (P6, P10, P12) mentioned \tool{} support learning by extending assistance beyond online settings. They expalned LearnMate is even helpful for planning the actual on-site class; context based planning/AI tool is also needed for actual on-site class. AI tool makes help more accessible, like a professor/being classes. whereas Gemini didn't help much completing the course.

One participant (P14), who teaches online courses, highlighted \tool{}'s potential to support instructors by summarizing course content and providing guidance that extends beyond what instructors can offer directly. They explained that \tool{} could save instructors time by generating review materials and offering embedded support for students, particularly in settings where instructors are unable to provide synchronous review sessions.
As P14 noted:
\textit{
``I teach an online course, and I think this assistant would be great for helping students review before quizzes or exams, especially since I don't always have time to do review sessions. Having AI support built directly into where the lectures are would be really helpful.''}

% One particpant (P14), as a online learning instuctor, mentioned \tool{} could help instructors through content summarization and guidance and give help beyond instutor. they expalined LM saves online learning instructors time by summarizing the content that will be taught and help students with parts that online instructors cannot help.

Four participants (P7, P9, P11, P13) emphasized that \tool{} helped learners better memorize content and prepare for exams through adaptive, real-time support. Participants described the system as particularly useful in courses with frequent grading, as it summarized key ideas and reinforced important concepts. They also noted that quiz-based adaptations reduced cognitive effort by adjusting plans based on performance, which supported learning even when learners were less motivated to actively optimize their study strategies. 
% Participants preferred detailed plans and structured learning materials, which they felt contributed to improved learning outcomes.
Participants preferred detailed plans and structured materials, which they felt improved their learning outcomes.
As P7 explained:
\textit{
``If I know exactly what I want to change, I can tell it and it can do that for me. But if I'm feeling lazy and don't want to think too much about how to optimize my learning, having the system adjust things based on a quiz or what I've learned before is easier and takes less effort than figuring everything out myself.''}

% Four pariticapent (P7, P9, P11, P13) mentioned enabling learners to better memorize content and prepare for exams through adaptive, real-time support. they expaliend ke it's really useful for those who talk a lot of grading and it helps you to summarize and give you all the key ideas of content . also Learners might be lazy and adjustment based on the quiz help them with less thinking. also Detailed plan and learning materials is preferred and will lead to higher score

\subsubsection{Learners Prefer Mixed-Initiative Planning over Fully Automated Learning}
Five participants (P2, P3, P4, P7, P13) expressed a preference for mixed-initiative planning over fully automated learning plans. Participants explained that due to skepticism toward AI accuracy and concerns about hallucinations or misinterpretation of personal needs, they were reluctant to fully delegate planning decisions to AI systems. Instead, they preferred an approach in which \tool{} provide structured prompts, initial drafts, and guidance, while learners retain control over planning decisions and adjustments.
Participants noted that fully automated plans could feel misaligned with their learning needs, whereas mixed-initiative planning allowed them to combine AI support with their own judgment and existing study habits. In particular, they described structured guidance—such as prompting for time availability and generating concrete schedules—as making plans feel more accountable and easier to follow, while still allowing learners to fine-tune plans based on personal preferences, performance, and evolving needs. This combination of structured AI guidance and human planning was perceived as more effective and trustworthy.
As P7 explained:
\textit{``I like being able to fine-tune things myself rather than having the system tell me exactly what to do. I don't have to follow everything the assistant gives me—I can use it as a rough outline and then make my own adjustments. Ultimately, I'm the one taking the quiz or exam, so I know what kinds of questions come up and what I need to focus on. Having that flexibility is really nice.''}

% Five partiicapned (P2, P3, P4, P7, P13) said given learners' skepticism toward AI accuracy and their existing planning habits, rather than fully automating planning, learners prefer mixed-initiative learning in which LMs provide structured prompts and guidance while learners retain control over planning and adjustment. Patcipanedt exaplined they are skeptical about whether AI can accurately interpret their needs and Hallusination will make the learning experience worse/mislead. But \tool{}  having specific time generated will make the adjusted plan more accountable, and makes learning more exaiser. Thus, combing structured guidance \tool{} provide with traditional human planning makes learning process easier and it's preferred. 

\subsubsection{Learning Support Systems Benefit from Richer Learner Input and Familiar LLM Interfaces}
Participants identified two key directions for improving future learning support systems. First, participants emphasized the importance of supporting richer learner input and diverse quiz formats to better capture learners' needs and understanding. Four participants (P2, P3, P8, P12) noted that learning support systems should support diverse quiz formats and richer learner input to better capture learners' preferences, goals, and understanding. Participants explained that providing background context—such as pacing preferences and learning goals—allowed the system to tailor plans more effectively and help learners get more value out of each learning session.
As P2 explained:
\textit{
``I liked the plan because it was easy to follow and more condensed in a way that made sense for taking a class. It felt like it understood that you want to get as much out of each session as possible and planned things accordingly, which was really helpful.''}

Second, participants highlighted the value of integrating learning support systems into familiar LLM interfaces to reduce learning overhead and support sustained engagement.
Four participants (P1, P9, P11, P15) suggested that learning support systems could be improved by integrating into familiar LLM interfaces. Participants explained that while \tool{} was useful, it required time to learn a new interface, whereas familiar tools such as Gemini already supported flexible, free-form interaction. They noted that reducing interface-switching and onboarding effort could increase long-term engagement and make learning support tools easier to adopt.
As P15 explained:
\textit{
``Gemini feels more free-form, and I'd be more likely to use it if it were trained on the material I was learning and let me interact with it directly. I liked that the learning tool could access the video and summarize it, but using a familiar interface would make it easier to use over time.''}

% Participants mentioned effective future learning support systems should imporve in two ways.
% Firts, four participants (P2, P3, P8, P12) mentioned they like \tool{} support diverse quiz formats and richer learner input to better capture learners' need and understanding. They descibe It is important for AI education tool to get background context of user perference/personalized so it gives as much out of each session as possible when learning.

% Additinoally, four participants (P1, P9, P11, P15) mentioned the system could improve by integrate into familiar LLM interfaces, as learners value flexible assessment, reduced learning overhead, and longer-term engagement to realize the system's benefits. They desbrie \tool{} as need learning time for the new tool, but once used to it will be better/useful

\section{Discussion}
%\subsection{Learning Experiences}
%\subsection{Learning Outcomes}

This work introduces \tool{} as an integrated learning workflow with three coordinated roles for LLMs: \texttt{PlanMate} scaffolds personalized, calendarized study plans; \texttt{StudyMate} provides real-time, course-grounded assistance with progressive disclosure; and \texttt{AdaptMate} enables user adaptation on learning plans based on formative assessment.
With \tool{} across studies, participants reported higher usefulness, ease, and satisfaction for real-time assistance and adaptive activities, higher SUS for the assistance component, and better learning outcomes on both quizzes.

The qualitative data explains why the system works as a \textit{workflow} rather than a one-off chatbot. Learners preferred course-verified answers they could expand on demand, appreciated real-time and accessible help (especially ESL learners), and relied on quiz-driven adaptation to surface unknown-unknowns and keep momentum when initial plans faltered. They also asked for durable artifacts and control: persistent calendars, provenance labels when widening beyond course materials, schedule-aware replanning, and reversible changes (\eg accept/modify/undo).

% Taken together, the findings point to three design commitments: personalization as process, the roles of LLMs in personalized learning support, and user control and agency for effective personalization. Below, we present the related implications.
Collectively, the findings point to three design commitments: personalization as an ongoing process rather than a one-time configuration, the strategic roles of LLMs in providing different types of personalized learning support throughout the learning workflow, and the critical importance of user control and agency for effective personalization.
% that builds rather than undermines learner autonomy.
% These commitments reflect not only technical capabilities but also pedagogical principles and interaction designs  about how learning occurs most effectively in digital environments. Below, we present the related implications that can guide the design of future personalized learning systems.
This requires not only technical capabilities for personalization and adaptation, but also interaction designs that help learners navigate complex content, maintain motivation, and develop self-regulation skills without excessive cognitive burden. 
Below, we outline design implications that can guide the design of future personalized learning systems.

\subsection{Personalization as Workflow: Building a Plan–Study–Adapt Loop}
Across both studies, \tool{} appears most effective when functioning as an orchestrated workflow where each component reinforces the others: structured planning through \texttt{PlanMate} reduces cold start and clarifies next learning steps; course-grounded, real-time support via \texttt{StudyMate} maintains learner engagement and attention while improving study efficiency; and quiz-driven adaptation in \texttt{AdaptMate} transforms assessment data into actionable plan modifications that sustain learner motivation and course completion. 
This integrated approach aligns with human-AI interaction (HAI) guidance to embed AI in existing routines with clear, user-visible feedback loops and handoffs \cite{amershi2019guidelines, lee2025veriplan, lee2026udefinedesigninguserworkflows}.

Our findings demonstrate that \tool{}'s effectiveness emerges from the synergistic integration of its three components rather than the sum of their individual contributions. The preliminary study revealed that isolated components (\texttt{PlanMate} or \texttt{StudyMate} alone) showed potential but produced limited measurable impact, while the full user study with integrated workflow yielded significant improvements in both learning outcomes and user experience.

\paragraph{\textbf{Design Implication:} Support learning as a continuous, adaptive process rather than discrete interactions.} 
% After each session, systems can present a brief ``session digest'' that summarizes what was studied and surfaces potential next steps, creating a clear handoff between planning, studying, and adaptation while keeping changes visible and revisable by the learner.
Our findings suggest that future systems could treat planning, studying, and adaptation as a single, evolving workflow state rather than separate features. 
After each session, systems can present a brief ``session digest'' that summarizes what was studied and highlights potential next steps, creating a clear handoff between planning, studying, and adaptation while keeping changes visible and modifiable by the learner.
This approach enables personalized learning systems to respond continuously and dynamically to actual progress rather than relying on predetermined predicted needs and isolated educational tasks common in existing systems \cite{maier2022personalized}, creating more effective and engaging learning experiences.

\subsection{Contextual and In-Situ Support for Lowering Help-Seeking Barriers among Diverse Learners}
Our findings from both studies demonstrate that contextual support, such as real-time assistance and domain-specific knowledge grounded in the study material, can lower help-seeking barriers for diverse learners and provide particular benefits for ESL learners. 
During the study, participants described \texttt{StudyMate}'s context-aware Q\&A as ``feeling like having an instructor next to me,'' noting that it reduced task-switching and was particularly helpful for ESL learners by enabling rapid, in-context definition and simplification.
Additionally, participants reported that placing the chatbot alongside course content within the same interface enabled in-situ support, substantially reducing the interaction overhead associated with navigating multiple platforms or waiting for instructor responses.
This immediate accessibility addresses critical gaps in online learning support, where traditional help-seeking mechanisms often fail due to delayed responses, lack of context, or intimidating formal processes. This aligns with existing work, which shows that actionable, audience-aware transparency improves sense-making and willingness to engage \cite{ehsan2021expanding}.

\paragraph{\textbf{Design Implication:} Combine contextual support with learning-suited, in-situ interfaces to lower help-seeking barriers.}
Future systems could integrate context-aware question interfaces that provide real-time assistance and domain-specific knowledge directly within learning-suited, in-situ interfaces, allowing learners to seek help without leaving their current learning context. 
Such designs can make help-seeking easier and more comfortable in a non-judgmental environment, reducing the social and logistical barriers that often prevent assistance-seeking in online learning.

% PM
\subsection{Structure over Blank Prompts: Plan Generation to Scaffolds Learners}
% Participants preferred structured starters and domain scaffolds to open-ended prompting, especially when they ``didn't know where to start.'' An immediate draft with a calendar-based plan lowered cognitive load and improved follow-through via chunked steps. This echoes existing work that highlights how non-expert users benefit from prompt scaffolds and explicit task parameters \cite{zamfirescu2023johnny}.
Both our findings demonstrate that structured guidance will make online learning more efficient by solving the cold start issue compared to blank prompts. 
Participants preferred structured starters and domain scaffolds to open-ended prompting, especially when they ``didn't know where to start.'' 
The immediate generation of calendar-based plans with time-chunked learning sessions reduced cognitive burden and enhanced plan execution through clear progression markers. This echoes existing work that highlights how non-expert users benefit from prompt scaffolds and explicit task parameters \cite{zamfirescu2023johnny}.

% \paragraph{\textbf{Design Implication.}} Our results indicate that future systems can reduce cold-start effort by offering user-friendly interface components (\eg buttons, sliders) for specifying personalized plan parameters (\ie goals, pace, time windows, path) that immediately render an editable plan and calendar, encouraging learners to refine rather than author from scratch.
\paragraph{\textbf{Design Implication:} Provide structured scaffolding that guides learning while maintaining user control.} 
Our results indicate that future systems can reduce cold-start effort by offering user-friendly interface components (\eg buttons, sliders) with structured guidance for specifying personalized plan parameters (\ie goals, pace, time, path) that immediately render an editable plan and calendar, encouraging learners to refine rather than author from scratch.
% This structured approach balances guidance with flexibility, allowing learners to customize by themselves while benefiting from expert-informed frameworks that reduce cognitive overhead.

% to do: starts from here
% SM
\subsection{Grounded First, Then Broad: Provenance-Aware Study Support}
% In our findings, course-grounded answers were more useful and time-efficient than generic LLM responses, yet some tasks benefited from broader resources. 
% Existing work also supports that progressive disclosure works best when paired with explicit provenance and scope so learners can choose depth and breadth intentionally \cite{rader2018explanations, ehsan2021expanding, lee2024ai}.
Our findings from both preliminary and full user studies revealed that course-grounded answers were consistently more useful and time-efficient than generic LLM responses. Findings from full user studies also indicated that the combination of contextual assistance and broader resources within limited length, followed by progressive disclosure mechanisms ensures both the depth and broadness of the support, and enable deeper exploration when desired.
Participants valued the context-aware answers and access to broader resources within limited length initially, followed by progressive disclosure mechanisms that enabled deeper exploration when desired.
Existing work also supports that progressive disclosure works best when paired with explicit provenance and scope so learners can choose depth and breadth intentionally \cite{rader2018explanations, ehsan2021expanding, lee2024ai}.

\paragraph{\textbf{Design Implication:} Prioritize course content, then limited external sources, then optional broader resources} 

Based on these findings, future systems could structure responses in layers: course-verified content with limited external sources as the default, expandable details through progressive disclosure, and optional toggles for broader curated sources labeled by provenance (\eg course-verified, external—curated, or low-confidence), enabling learners to consciously broaden scope.

% AM
\subsection{Formative Assessment as a Driver for Transparent, Learner-Steerable Adaptation}
Findings from the final user study demonstrate that quiz-driven formative assessment is essential for revealing knowledge gaps that learners cannot easily identify on their own, providing both a clear justification and an actionable starting point for adapting learning plans. 
Participants reported greater confidence in plan adaptation when they could see explicit connections between their quiz performance and the resulting plan changes, resonating with explanatory debugging principles that emphasize linking system modifications to observable user evidence \cite{kulesza2015principles}. 
This also aligns with prior work on conversational learning assessment systems, which leverage performance signals to guide targeted educational interventions (\eg QuizBot) \cite{ruan2019quizbot, BANIHASHEM2022100489}.

At the same time, our findings highlight that effective adaptation in learning contexts must preserve learner agency. 
Participants described motivation dips over time and expressed concerns about becoming over-reliant on automated guidance. 
They particularly valued schedule-aware replanning and reversible changes, noting that the ability to accept, modify, or reject proposed adaptations helped them maintain a sense of ownership over their learning process. 
These preferences align with prior research emphasizing the importance of user agency for supporting metacognitive development and self-regulated learning \cite{deschenes2020recommender, brod2023agency}, as well as HAI guidelines that caution against over-automation in high-stakes tasks \cite{amershi2019guidelines}, particularly given recent evidence that agentic systems are prone to failure on long-horizon tasks \cite{wang2026longhorizontaskmiragediagnosing}.

\paragraph{\textbf{Design Implication:} Use formative assessment to drive transparent, learner-controllable adaptation.}
Our findings suggest that future learning systems should use formative assessment signals to propose explainable plan adaptations while keeping learners in control of final decisions. 
Systems could present clear rationales linking performance to suggested changes (\eg ``missed \textit{X} $\Rightarrow$ add practice \textit{Y}; slow on \textit{Z} $\Rightarrow$ move \textit{W} to tomorrow''), alongside explicit controls to accept, modify, or undo each update. 
This approach enables actionable, low-effort plan updates that allow learners to make informed adjustments without the burden of re-planning from scratch. 
These findings highlight the value of mixed-initiative designs in which AI provides evidence and structured entry points for adaptation while learners retain decision authority---an approach that is particularly well-suited for high-stakes learning contexts, where trust, accountability, and self-regulation are critical.

At the same time, our findings suggest that scaffolding in AI-supported learning systems should not remain static over time. While some supports may need to persist, such as course-grounded answers in \texttt{StudyMate}, others may be better designed to fade as learner competence grows, such as step-by-step planning prompts in \texttt{PlanMate} or highly directive adaptation suggestions in \texttt{AdaptMate}. Otherwise, these scaffolds risk offloading metacognitive work that learners need to develop for themselves, including self-assessment, planning, and replanning. Future systems can therefore explore adaptive fading strategies that gradually shift responsibility back to learners as they gain familiarity and confidence, drawing on scaffolding fading principles in educational research~\cite{collins1991cognitive, dennen2013cognitive}.

\section{Limitations \& Future Work}
While \tool{} demonstrates potential for improving learning outcomes for online learning, several limitations should be acknowledged and addressed in future research. 

% {\color{blue}{
% First, while the chosen baseline (Khan Academy with Gemini-2.5-pro) reflects how students commonly supplement online learning in practice, more integrated AI-assisted learning experiences have since emerged that combine course platforms with built-in LLM support (\eg KhanMigo\footnote{\url{https://www.khanmigo.ai/}}). 
% % These tools may have addressed some of the friction participants in our baseline condition experienced, such as tab-switching between platforms. 
% Future work could compare \tool{} against such systems to better isolate the specific benefits of its closed-loop planning, studying, and adaptation workflow.}}
First, while the chosen baseline (Khan Academy with Gemini-2.5-pro) reflects how students commonly supplement online learning in practice, more integrated AI-assisted learning experiences have since emerged that combine course platforms with built-in LLM support (\eg KhanMigo\footnote{\url{https://www.khanmigo.ai/}}). 
% These tools may have addressed some of the friction participants in our baseline condition experienced, such as tab-switching between platforms. 
Future work could compare \tool{} against such systems to better isolate the specific benefits of its closed-loop planning, studying, and adaptation workflow.

Second, our evaluation focused exclusively on Khan Academy's World History Project to maintain consistency for comparative analysis purposes. This domain-specific focus may restrict the generalizability of findings. Future work can expand to diverse academic domains, including STEM fields (mathematics, physics), engineering, and other subjects, to assess the system's effectiveness across different knowledge types and learning objectives.
Similarly, our empirical evaluation is limited to data from 40 participants (24 in the preliminary study, 16 in the final study). While our within-subjects design provides adequate statistical power for detecting significant effects, the limited sample size and demographic diversity may restrict the generalizability of findings. Future work can explore larger-scale evaluations with more diverse participant populations to more conclusively validate the outcomes presented in this paper and build on the results to highlight additional insights. 
% {{\color{blue}{Moreover, our evaluation relies on learner-reported perceptions and quiz performance rather than expert assessment of output quality. Future work could include blind expert evaluation of \tool{}'s generated plans and responses relative to general-purpose LLMs to better assess pedagogical efficacy independently of learner preferences.}}
Moreover, our evaluation relies on learner-reported perceptions and quiz performance rather than expert assessment of output quality. Future work could include blind expert evaluation of \tool{}'s generated plans and responses relative to general-purpose LLMs to better assess pedagogical efficacy independently of learner preferences.

In addition, \tool{} currently focuses on the planning, studying, and adaptation components of the learning workflow, but this does not encompass the complete online learning experience. Participant mentioned that reminder notifications are also important to help them stay on track and maintain motivation over time. 
Future work could investigate what and how proactive notification systems and motivational interventions can enhance long-term engagement and help learners persist through course completion in online learning environments. 
% Additionally, our study focused on short-term learning outcomes, leaving the long-term impact on skill retention and transfer unclear.
% {\color{blue}{Furthermore, the short study duration limits our ability to fully evaluate \texttt{StudyMate} and \texttt{AdaptMate}, both of which are designed for longer-term use. 
% % \texttt{StudyMate} accumulates cross-session learning history, and \texttt{AdaptMate} proposes plan revisions that become most meaningful over multiple sessions. 
% A single-session study cannot capture these longitudinal dynamics, nor can it assess whether learners develop genuine self-regulation skills or grow increasingly reliant on system scaffolding over time. Future work should evaluate \tool{} over multiple weeks in naturalistic settings to better understand long-term learning outcomes, skill retention, and changes in learner autonomy.}}
Furthermore, the short study duration limits our ability to fully evaluate \texttt{StudyMate} and \texttt{AdaptMate}, both of which are designed for longer-term use. 
% \texttt{StudyMate} accumulates cross-session learning history, and \texttt{AdaptMate} proposes plan revisions that become most meaningful over multiple sessions. 
A single-session study cannot capture these longitudinal dynamics, nor can it assess whether learners develop genuine self-regulation skills or grow increasingly reliant on system scaffolding over time. Future work should evaluate \tool{} over multiple weeks in naturalistic settings to better understand long-term learning outcomes, skill retention, and changes in learner autonomy.

Finally, future work could involve the expansion of \tool{}. The system can be adapted to support underrepresented learner populations such as students with ADHD, dyslexia, or those with varying levels of digital literacy who may benefit from different interaction modalities or interface designs. 
Additionally, \tool{} can be developed as a flexible plugin that integrates with existing online learning platforms such as Khan Academy, MOOCs, and YouTube tutorials to maximize accessibility and impact. This approach would provide users with personalized learning support within familiar environments, reducing barriers to adoption and ensuring that diverse learners can benefit from tailored educational assistance. 
Addressing these limitations could further strengthen \tool{}'s usability and effectiveness, making it a more adaptable and scalable solution for online learning.

\section{Conclusion}
This study introduces \tool{}, an LLM-powered system designed to support learners in online learning through personalized study plans, real-time contextual assistance, and adaptive learning activities that adjust based on ongoing progress assessment. By leveraging natural language processing and interactive prompts, \tool{} enhances the online learning experience, extending the effectiveness and accessibility of online education for diverse learner populations.

Our findings from both preliminary and final user studies demonstrate that \tool{} significantly improves learning outcomes and user experience compared to traditional online learning approaches. 
Each component provides unique benefits: personalized study plans improve accountability and ease of following through structured guidance; real-time contextual assistance enhances accessibility and learning effectiveness, particularly for memorization-intensive subjects; and adaptive learning activities improve learning confidence and autonomy by identifying strengths and weaknesses that learners cannot recognize independently. 
Findings from final user studies also show that when integrated as a complete workflow, \tool{} demonstrates that the orchestrated combination of planning, studying, and adaptation creates a synergistic effect that enhances overall learning effectiveness, efficiency, engagement, and motivation beyond what individual components could achieve in isolation. 
Additionally, the study highlights the importance of integrating LLMs in educational contexts to provide instructor-like support in online settings where traditional guidance is limited.

The insights gained from the design and evaluation of \tool{} provide valuable guidelines for developing future LLM-powered educational tools that enhance learning outcomes through personalized workflows and adaptive support. 
Our design implications emphasize building plan-study-adapt feedback loops, offering provenance-aware support, and maintaining user agency through transparent system adaptations. 
Future work will explore expanding the system's capabilities and examining its application in broader educational contexts.
Together, we aim to foster a more effective, accessible, and learner-centered online learning experience.

% out findsings from both prelimary and finauser styudy shows that planning can omprove xxx, study can improce xxx, apt can impreove xxx. while findings from and compare of prelimary and finauser styudy shows the system works as a whole can improve xx(all can be find in findings).
% we present a system that xxxx. xxxx. xxx. xxxx(as in discssino) and also future worj. 
% Through these efforts, we seek to create a more effective, accessible, and user-centered online learning experience.

\begin{acks}
This work was supported by the Sheldon B. and Marianne S. Lubar Professorship.
\end{acks}

%----------------------------------------------
% \subsection{Formative User Study}
% we don't want to build gemini
% we want to know how to personalized learning

% \subsection{Final User Study}

% \section{Results}
% % explain more concrete
% % conditions: provided with same syllabus

% \section{Discussion}

%% the bibliography file.
%\balance
\bibliographystyle{ACM-Reference-Format}
% \balance
\bibliography{bibliography}

\end{document}